\documentclass{sig-alternate}
\pdfoutput=1
\usepackage{latexsym, mathrsfs}
\usepackage{graphicx}
\usepackage{soul}
\usepackage[T1]{fontenc}
\usepackage[usenames,dvipsnames,svgnames,table]{xcolor}
\usepackage[linesnumbered,ruled,procnumbered]{algorithm2e}
\usepackage{times}
\usepackage{paralist}
\usepackage{multirow}
\usepackage{xspace}
\usepackage{url}
\usepackage[caption=false]{subfig}
\usepackage{hhline}
\usepackage{enumitem}
\usepackage{balance}
\usepackage{amsmath}
\usepackage{array}

\newcolumntype{K}[1]{>{\centering\arraybackslash}p{#1}}
\newcommand{\stitle}[1]{\vspace{0.5em}\noindent\textbf{#1}}

\numberwithin{equation}{section}

\newtheorem{lemma}{Lemma}

\newtheorem{example}{Example}

\newcommand{\resolved}[1]{\textcolor{green}{[Amol: #1]}}
\renewcommand{\resolved}[1]{}

\newcommand{\ds}{{RStore}\xspace}

\newcommand{\bt}{{\sc Bottom-Up}\xspace}
\newcommand{\sh}{{\sc Shingle}\xspace}
\newcommand{\dfs}{{\sc DepthFirst}\xspace} 
\newcommand{\bfs}{{\sc BreadthFirst}\xspace}
\newcommand{\deltat}{{\sc Delta}\xspace}
\newcommand{\subc}{{\sc SubChunk}\xspace}
\newcommand{\compr}{{\sc Comp. Ratio}\xspace}

\newcommand{\pp}{\mathcal{P}\xspace}
\newcommand{\mm}{\mathcal{M}\xspace}

\newcommand{\eat}[1]{}

\newcommand{\techreport}[1]{}

\newcommand{\techreporttext}[1]{}

\newenvironment{denselist}{
    \begin{list}{\small{$\bullet$}}%
    {\setlength{\itemsep}{0ex} \setlength{\topsep}{0ex} 
    \setlength{\parsep}{0pt} \setlength{\itemindent}{0pt}
    \setlength{\leftmargin}{1.2em}
    \setlength{\partopsep}{0pt}}}%
    {\end{list}}

\newcommand{\topic}[1]{\vspace{-3.5pt}\smallskip \smallskip \noindent{\bf #1:}}

\begin{document}
\title{RStore: A Distributed Multi-version Document Store}

\numberofauthors{1}
\author{
\alignauthor Souvik Bhattacherjee, Amol Deshpande\\
\affaddr{University of Maryland, College Park} \\
\email{\{bsouvik, amol\}@cs.umd.edu}
}

\maketitle


\begin{abstract}
We address the problem of compactly storing a large number of versions (snapshots) of a collection of {\em keyed documents} or {\em records} in a distributed environment, while 
efficiently answering a variety of {\em retrieval} queries over those, including retrieving full or partial versions, and evolution histories for specific keys. 
We motivate the increasing need for such a system in a variety of application domains, carefully explore the design space for building such a system and the various storage-computation-retrieval trade-offs, and discuss how different storage layouts influence those trade-offs. We propose a novel system architecture that satisfies the key desiderata for such a system, and offers simple tuning knobs that allow adapting to a specific data and query workload.
Our system is intended to act as a layer on top of a distributed key-value store that houses the raw data as well as any indexes. We design novel off-line storage layout algorithms for efficiently partitioning the data to minimize the storage costs while keeping the retrieval costs low. We also present an online algorithm to handle new versions being added to system.
Using extensive experiments on large datasets, we demonstrate that our system operates at the scale required in most practical scenarios and often outperforms standard baselines, including a delta-based storage engine, by orders-of-magnitude.

\end{abstract}

\section{Introduction}

The desire to derive valuable insights from large and diverse datasets produced
in nearly all application domains today, has
led to large collaborative efforts, often spanning multiple organizations.
The iterative and exploratory nature of the data science process, combined with an increasing
need to support debugging, historical queries, auditing, provenance, and reproducibility, means that a large 
number of {\em versions} of a dataset may need to stored and queried. This realization has led to 
many efforts at building data management systems that support versioning as a first-class construct,
both in academia~\cite{buneman2004,BhattacherjeeCH15,MaddoxGEMPD16,ground} and in industry (e.g., git, Datomic, noms).
Unlike archival storage systems which also maintain large histories, 
these systems typically support rich versioning/branching functionality and, in some cases, complex queries
over versioned information. 


We motivate the design and development of our system using a concrete example from a real-life scenario.
\begin{example}
	A healthcare provider who wants to perform different types of diagnostic and prognostic analytics may need to continuously maintain and analyze Electronic Health Records (EHRs) of thousands to millions of patients. The EHR dataset is continuously changing through addition/deletion of new patient EHRs and updates to existing ones. For many practical reasons, results of applying any analytics are usually stored in the same EHR documents. Data analysts usually target a particular group of people when running analytical tasks in order to minimize the number of variables, e.g., people between age 50 - 60, belonging to a given ethnicity, with certain other characteristics, etc. As a result the number of updates per version usually remains restricted to a small percentage w.r.t the total pool of patients. Different teams of data scientists, with different goals, may be tweaking, training, and applying predictive models to those documents at the same time. Because of decentralized nature of the updates and increased use of collaborative analytics, the resulting version histories are mostly ``branched". For accountability and debugging, it is essential that the precise details and provenance of all of those steps are maintained; e.g., an analyst must be able to clearly identify which versions of the EHRs were used to train a particular model, or which models were used to derive a specific individual prediction. 
It is also necessary for them to retrieve all or a subset of past versions of patients to analyze them for insights. Further, looking up a patient history from the point it enters their system is a very common query for them. 
The EHR schemas also evolve continuously when new data points that correspond to non-existing attributes are added in the form of new medical tests or measurements to a subset of the EHRs. Given the scale of the data, continuously evolving and semi-structured schema, and a desire to support distributed collaboration, key-value stores are often a natural option for storing such data (an extraction step to convert from the highly normalized relational databases where the original data is stored is quite common).
\end{example}

Similar requirements are beginning to arise in diverse application domains such as knowledge bases, content management systems, computation biology, and many others. Although there has been much work on version control systems in recent years, none of those prior systems are designed for hosting 
versions of a collection of keyed records or 
documents in a distributed environment or a cloud, while providing querying functionality similar to
the wildly popular {\em key-value stores}. Key-value stores, a term loosely used here to describe any SQL/NoSQL system that supports key-based retrieval~\cite{Cattell2010} (e.g., Apache Cassandra, HBase, MongoDB) 
are appealing in many collaborative scenarios spanning geographically distributed teams,
since they offer centralized hosting of the data, are resilient to failures, can easily scale out, and can 
handle a large number of queries efficiently. However, those 
do not offer rich versioning and branching functionality akin to hosted version control systems (VCS) like {\em GitHub}. 
The necessity of maintaining document versions have resulted in several quick and dirty extensions of systems like MongoDB and Couchbase, to satisfy immediate user needs~\cite{mongodb-blog, couchdb-blog}. Unfortunately, the solutions presented there have several limitations and fail to provide any guarantees on the quality of the solution.  

In this paper, our primary focus is to provide versioning and branching support for collections of records with unique identifiers that can act 
as {\em primary keys}. Like popular NoSQL systems, we aim to support a flexible data model; records with varying sizes, ranging from a few bytes
to a few MBs; and a variety of retrieval queries to cover a wide range of use cases. Specifically, similar to NoSQL systems, we aim to support 
efficient retrieval of a specific record in a specific version (given a key and a version identifier), or the entire evolution history for a given 
key. Similar to VCS, we aim to support retrieving all the records belonging to a specific version to support use cases
that require updating a large number of the records (e.g., by applying a data cleaning step). Finally, since retrieval of an entire version might
be unnecessary and expensive, we also aim to support {\em partial} version retrieval given a range of keys and a version identifier.
In addition, we aim to support efficient {\em ingest} (``commit'') of new versions from users, where the change from the previous version 
(``delta'') may be a small update to one record, or updates to a large subset of the records.

We begin with a careful exploration of the design space, outline the different trade-offs, and discuss the limitations of the baseline alternatives with respect to the desired requirements listed above. As observed in prior work (e.g.,~\cite{BhattacherjeeCH15}), there is a natural trade-off between the storage requirements and the querying efficiency. However, the baseline approaches suffer from more fundamental limitations. (a) First, most of those approaches cannot directly support point queries targetting a specific record in a specific version (and by extension, full or partial version retrieval queries), without constructing and maintaining explicit indexes. (b) Second, all the viable baselines fundamentally require too many back-and-forths between the retrieval module and the backend key-value store; this is because the desired set of records cannot be succintly described. (c) Third, ingest of new versions is difficult for most of the baseline approaches. (d) Finally, exploiting ``record-level compression'' is difficult or impossible in those approaches; this is crucial to be able to handle common use cases where large records (e.g., documents) are updated frequently with relatively small changes.

To address these problems, we investigate a new architecture 
that partitions the distinct records into approximately equal-sized ``chunks'', with the goal to minimize the number of chunks that need to be retrieved for a given query workload. We show how the system can adapt to different data and workload requirements through a few simple tuning knobs. The key computational challenge boils down to deciding how to optimally partition the records into chunks; we draw connections to well-studied problems like compressing bipartitite graphs and hypergraph partitioning to show that the problem is NP-Hard in general. We then present a novel algorithm, that exploits the structure of the version graph, to find an effective partitioning of the records. We have built a working prototype of our system, called \ds, on top of the Apache Cassandra key-value store. \ds can handle arbitrary types of records, including semi-structured (JSON/XML) documents, and text or binary files. We conduct an extensive experimental evaluation over a large number of synthetically constructed datasets to show the effectiveness of our system and to validate our design decisions.

Our key contributions can be summarized as follows: (1) We systematically explore the design space for supporting versioning as a first-class construct in distributed key-value stores; (2) We present a detailed analysis of the different trade-offs and how different baselines fare with respect to those; (3) We propose a flexible system architecture that supports the key desiderata through use of ``chunking''; (4) We design novel partitioning algorithms that exploit how the versions relate to each other to identify good chunking strategies; 
(5) We present an online algorithm to keep the partitioning and the indexes up-to-date as new versions are committed. 
(6) We have built a working prototype on top of the Apache Cassandra key-value store, which we use to validate our design decisions. We expect that \ds, like many NoSQL stores, will primarily be deployed in a distributed environment; however, it can also be used in a local cluster.

\vspace{-10pt}
\section{System Design}

We start with a brief description of the underlying data model, followed by a description of the retrieval queries we aim to support. Thereafter, we 
provide a brief overview of the overall system architecture.

\subsection{Data and Query Model}\label{ssec:data_n_query_model}

\stitle{Data Model.} The primary unit of storage and retrieval in our system is a {\bf record}, which may refer to a tuple/row in a tabular dataset, a JSON document in a document collection, or a time series. A record is considered to be {\em immutable}, and any change to it results in a new {\em version} of the record. We make no assumptions about the structure, type or the size of a record, except for assuming the existence of a {\bf primary key}, denoted $K_i$, that can be used to uniquely identify a specific record within a collection of records. For simplicity, we assume there is a single such collection (also called a {\em dataset}) that the system needs to manage, that is being parallely modified by a team of users over time in a collaborative fashion, resulting in a set of {\bf versions} over time. We assume there is a single {\em root} version of the dataset, from which all other versions are derived (an empty root version can be added to handle the scenario where there are multiple starting collections).  

Let $V = \{V_0, V_1, \ldots, V_{n-1}\}$ denote the set of {\em versions} stored in the system; each version is identified uniquely by a {\bf version-id} (either an auto-incremented value, or {\em hashes} as in {\em git}). 
A new version is derived from an existing version through an update operation or a transformation, that essentially boils down to modifying/deleting existing records and/or adding a new set of records. We denote the set of changes from $V_i$ to $V_j$ by $\Delta_{i,j}$ and refer to it as the {\em delta} from $V_i$ to $V_j$. Note that in this case, $\Delta_{i,j}$ is symmetric, i.e., $\Delta_{i, j}$ may be used to derive $V_i$ from $V_j$ as well, thus making $\Delta_{i, j} = \Delta_{j, i}$. These derivations are encoded in the form of a directed {\em version graph}. 



\stitle{Composite Keys.}
Since a record may be unchanged from one version to the next, to be able to refer to a specific record within a specific version, we use a {\bf composite key}: $\langle$primarykey, version-id$\rangle$, where the second part refers to the {\bf version-id} of the version where the record was created. This allows us to uniquely reference records within a global address space. We chose to use {\bf version-id} of the appropriate version instead of an auto-incremented value as the latter introduces additional synchronization overhead in a decentralized setting with no obvious benefits.

\stitle{Query Model.} In a collaborative setting with large datasets, the query workload may consist of a variety of queries, with differing characteristics.
\begin{list}{$\bullet$}{\leftmargin 0.15in \topsep 1pt \itemsep 1pt}
\item {\em Record Retrieval:} Analagous to a key-value store, a user/application may want to retrieve a record with a specific primary key $K$ from a specific version $V$. Note that we cannot simply retrieve the record with the composite key $\langle K, V \rangle$, since the record may have originated in one of the predecessor versions to $V$. \ul{This, in fact, forms a major challenge in this setting.}
 \item {\em Version Retrieval: } Analogous to typical VCS, here the goal is to retrieve the entire version given a version-id, i.e., all the records that belong to the version.
 \item {\em Range Retrieval: } This query enables retrieving a version partially, by specifying a range of primary keys and a version-id. 
 \item {\em Record Evolution:} Finally, we may want to analyze the evolution of a record from its point of origin to the current state of the system. In other words, given a {\em primary key}, we want to find all the different records with that primary key across all versions.
\end{list}


\begin{figure}[t]
\vspace{-5pt}
\includegraphics[trim=5cm 10.5cm 4cm 10.5cm, width=3.5in]{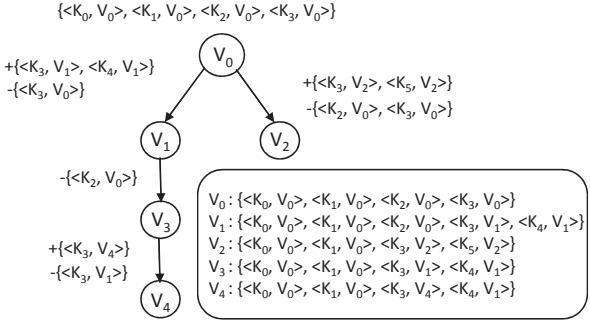}
\vspace{-29pt}
\caption{An Example Version Graph with 5 Versions}
\vspace{-14pt}
\label{fig:version_graph}
\end{figure}

\vspace{-5pt}
\begin{example}\label{ex:composite_key}
Fig.~\ref{fig:version_graph} displays a version graph with five versions $V_0\ (root), V_1, V_2, V_3, V_4$, 
with a total of nine {\em distinct} records. 
We create composite keys for the records in $V_0$ by adding $V_0$ as the second component to the keys.
$V_1$ is derived by modifying $K_3$ of $V_0$ and adding a new record $\langle K_4, V_1\rangle$. In this case $\Delta_{0,1} = \{+\langle K_3, V_1 \rangle, +\langle K_4, V_1\rangle, -\langle K_3, V_0\rangle\}$. 
$V_2$ is derived from $V_0$ (and after $V_1$) by modifying $K_3$ as well, adding a new record $\langle K_5, V_2\rangle$, and deleting record $\langle K_2, V_0\rangle$. 
$V_3$ is derived next by deleting record $\langle K_2, V_0\rangle$ from $V_1$. Finally, $V_4$ is derived by modifying $\langle K_3, V_1\rangle$ of $V_2$. Note that the derived version forms the version identifier component in the composite key, which is also the version in which the particular record appears for the first time. 
To retrieve a specific record, say $K_3$ from version $V_3$, it is not sufficient to look for composite key $\langle K_3, V_3 \rangle$ (which does not exist), rather, we need to maintain a version-to-record mapping (Fig.~\ref{fig:version_graph}), that must be consulted to identify the composite key to be retrieved ($\langle K_3, V_1 \rangle$ in this case).
\end{example}

\subsection{Key Trade-Offs}\label{ssec:tradeoff}
We begin with a brief discussion of the key trade-offs in storing such versioned datasets in the cloud, and then evaluate 3 baseline options with respect to those trade-offs.
\begin{list}{$\bullet$}{\leftmargin 0.12in \topsep 0pt \itemsep -1pt}

\item {\bf Storage and compression.} There are two considerations here. First, we would prefer to only store a single copy of a record that appears in multiple versions. This however complicates performance of full or partial version retrieval queries since the requisite records may be stored all over the place. Second, we would like \ds to handle records of varying sizes, from a few bytes to a few MBs. In the latter case, there may be small differences between two different versions of a record (e.g., only a single attribute may be updated in a large JSON document). One way to exploit this overlap is to store the two versions of the record together in a ``compressed'' fashion, with specific compression technique chosen according to the data properties (e.g., one may store ``deltas'' (differences) between the two records, or use an off-the-shelf compression tool that in effect does the same thing). Such compression, however, negatively impacts the query performance by restricting the data placement opportunities.


\item {\bf Query performance.} Different partitioning and layout schemes are appropriate for the different classes of queries listed above. Record evolution queries are best served by grouping together all the different records with the same primary key, whereas full version retrieval queries prefer grouping together all records that belong to the same version. A general-purpose system must offer knobs that allow adapting to a specific query workload.
\item {\bf Online updates.} Another important consideration is the ability to handle updates, i.e., new versions being added. Ideally the cost of incorporating a new version is proportional to the size of the update itself, i.e., the difference between the new version and the version it derives from. 
\end{list}

\noindent
Next, we discuss a few baseline approaches that serve as layers on top of a key-value store, and how they fare w.r.t. these trade-offs.
\begin{list}{$\bullet$}{\leftmargin 0.12in \topsep 0pt \itemsep -1pt}
\item  \underline{\bf Single address space:} Perhaps the simplest option is to store the records directly, using the composite key as the key for the underlying key-value store. Although simple to implement and offering best performance for updates (ingest), this approach has several disadvantages. First, there is no way to use compression to reduce storage requirements, since different records with the same primary key are stored separately. Second, given a specific version $V$ and a specific primary key $K$, retrieving the record with that primary key from that version (if present) requires an additional index. This is because of the way composite keys are generated -- we first need to identify the predecessor version to $V$ where that primary key was last modified. This complicates the execution of all the queries listed above. Not only does the index have to be repeatedly consulted, we may need to issue many queries against the backend key-value store.
\item \underline{\bf ``Sub-chunk'' approach:} Here, we group together all the records with the same primary key $K$, and store it in compressed fashion using $K$ as the key; we call such a group of records with the same primary key a {\bf sub-chunk}. This approach has the best storage cost and best performance for record evolution queries (and possibly single record queries, if the average number of different records per primary key is small). However, full or partial version retrieval queries require retrieving significant amounts of irrelevant data, especially if the data is not highly compressible (i.e., different records with the same primary key are more different than similar). 
Further, ingest is expensive since each of the relevant sub-chunks must be retrieved, de-compressed, and compressed after adding the appropriate record. \\
One alternative here is to create multiple sub-chunks per primary key, which results in retrieving less data and also speeds up online updates (the ``single address space'' approach is a special case). However, this negates much of the simplicity of the approach, since additional indexes need to be used to identify the specific sub-chunks that contains the data for a given version.
\item \underline{\bf Delta approach:} Here, analogous to how version control systems like {\em git} work, for each version, we store the difference from its predecessor version, i.e., the ``delta'' that allows us to get to the version from the predecessor version. The predecessor version itself may be stored as a delta from its predecessor and so on, forming {\em delta chains}. The main advantage of this approach is that updates are easy to handle, especially since we assume that a new version is presented as a delta from its predecessor version. Assuming that the delta is computed by exploiting similarities at the level of records, this approach naturally accrues the benefits of compression. However, performance of key-centric queries, i.e., specific record queries and record evaluation queries, is very poor for this approach. Even partial retrieval queries are difficult to do with this approach.
\end{list}
Table~\ref{table:baseline-alg} summarizes some of these trade-offs, by showing expressions for various different costs under some simplifying assumptions. Specifically, we assume a version with $m_v$ records, with a sequence of changes each updating a fraction $d$ of the records; thus the version graph here is a ``chain''. Note that this is a worst-case scenario for the delta approach; however, the main problem with the delta approach is partial or single record retrieval queries, where it has abysmal performance. 

\subsection{Too Many Queries Problem}\label{ssec:toomanyq}
None of the baseline approaches are thus appropriate for storing and querying a large number of versions of keyed records. Further, all of these approaches require {\em making a large number of queries} to the underlying key-value store for full or partial version retrieval. This is because the records belonging to a specific version $V$ cannot be easily described. For example, in the first approach (and the partial sub-chunk approach), we need to use separate indexes to identify the ``keys'' that must be retrieved, and all of those must be retrieved separately from each other (efficient support for large IN queries from the key-value store may help, but only shifts the problem to the key-value store). Similarly, in the Delta approach, all the requisite deltas must be retrieved one-by-one. 

To validate our claim, we performed a simple experiment using Apache Cassandra. Each version in the dataset has about 100K 100-byte records, with a total of 1 million unique records stored in the KVS. The query here is to reconstruct a version, i.e., we need to retrieve around 100K records for every version reconstruction query from the KVS. In the naive setting, we maintain a chunk of unit size and issue around 100K requests to the KVS. In comparison, if we create larger sized chunks using a random assignment of records to chunks, we need to retrieve more number of chunks than exactly required to recreate a version. However the overhead of retrieving additional chunks and scanning through them to extract the records is significantly less. This illustrates the significant benefits of reducing the number of queries made to the key-value store. Unfortunately, because of the aforementioned problem, this problem must be solved by explicitly creating ``chunks'' of records, where records belonging to the same set of versions are grouped together.

\eat{
\begin{table}[h]
\centering
 \begin{tabular}{| c | c |}
  \hline
  Chunk size & Time (in secs.) \\ \hline
  1 & 65.415 \\ \hline
  10 & 14.175 \\ \hline
  100 & 3.098 \\ \hline
  1000 & 1.072 \\ \hline
  10000 & 0.562 \\
  \hline
 \end{tabular}
 \caption{\small Benefits of ``chunking''}
\label{table:chunk-exp}
\end{table}
}

\begin{table}[h]
\centering
\begin{tabular}{|c|c|c|c|c|c|}
 \hline
 Chunk size & 1 & 10 & 100 & 1000 & 10000 \\ \hline
 Time (in secs.) & 65.42 & 14.18 & 3.10 & 1.07 & 0.56 \\ \hline
\end{tabular}
\label{table:chunk-exp}
\end{table}

\begin{table*}
\centering
{\small
 \begin{tabular}{| c | c | c | c |}
  \hline
  Algorithms & Storage Space    & Random Version (total data, \#queries)  & Point Query \\ \hline
  Independent w/chunking & $nm_{v}s$            & $m_{v}s,~~ {m_{v}s}/{s_c}$ & $s_c,~~1$ \\
  \deltat    & $m_{v}s + cd(n-1)m_{v}s$ & $m_{v}s+{cd(n-1)m_{v}s}/{2}, ~~n/2$ & $m_{v}s + {cd(n-1)m_{v}s}/{2}, ~~n/2$\\
  \subc & $m_{v}s + cd(n-1)m_{v}s$ & ${m_{v}(s + cd(n-1)s)}, ~~m_{v}$ & $s + cd(n-1)s, ~~1$ \\
  Single-address space & $m_{v}s + d(n-1)m_{v}s$ & $m_{v}s, ~~m_{v}s$ & $s, ~~1$ \\
  \hline
 \end{tabular}
  }
 \caption{\small Comparing the different options for storing versioned records along different dimensions under some simplifying assumptions. $n$ = number of versions (arranged in a chain); $m_v$ = number of records in a version (constant), $d$ = fraction of records that are updated in every version update, $c$ = compression ratio (typically $c, d \ll 1$), $s$ = size of a record, $s_c$ = size of a chunk. For the queries, the table shows: amount of data retrieved, number of queries. This analysis assumes the cost of consulting any indexes is negligible.}
\label{table:baseline-alg}
\end{table*}

\subsection{Architecture}
Figure~\ref{fig:sysarch} shows the high-level architecture of our system. In what follows, we describe the primary components that constitute our system, namely (i) Data Ingest Module, (ii) Data Placement Module, and (iii) Query Processing Module, as well as the different design choices that were made while building this system. 

\eat{
\begin{figure}[t]
\includegraphics[trim=2cm 8cm 8.2cm 4.7cm, width=3in,clip]{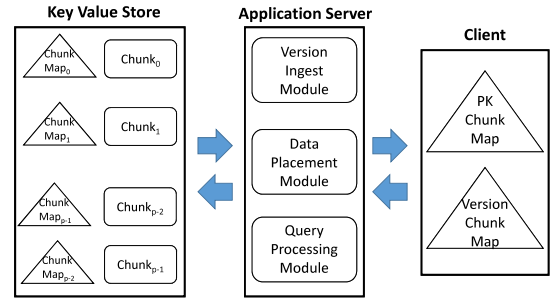}
\vspace{-12pt}
\caption{System Architecture}
\vspace{-12pt}
\label{fig:sysarch}
\end{figure}
}

\begin{figure}[t]
\includegraphics[trim=2cm 9.5cm 2.5cm 9.5cm, width=3in, clip]{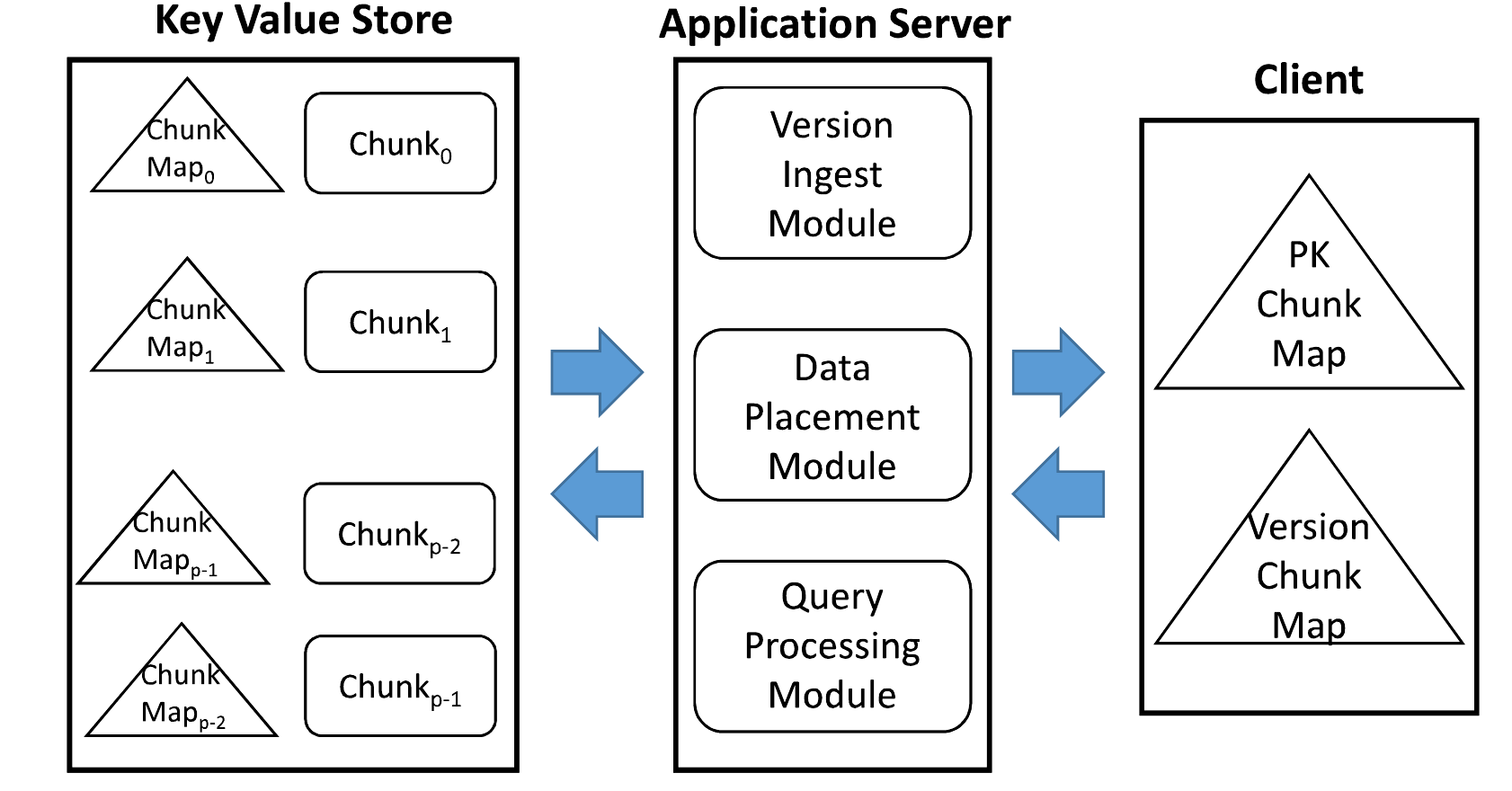}
\vspace{-12pt}
\caption{\small System Architecture}
\vspace{-12pt}
\label{fig:sysarch}
\end{figure}

\topic{Backend Key-value Store}
Our system is intended to act as a layer on an extant distributed key-value store, in order to leverage the significant research and implementation that has gone into designing scalable, fault-tolerant systems. Our implementation specifically builds on top of Apache Cassandra, but we only assume basic {\tt get/put} functionality from it. As shown in Figure \ref{fig:sysarch}, the basic unit of storage in the key-value store a {\em chunk} of records, with the keys called {\em chunk-ids}; chunk-ids are generated internally and are not intended to be semantically meaningful. Each chunk is divided into sub-chunks, each of which corresponds to records with the same primary key and are stored in a compressed fashion; sub-chunks often may contain only one record. In addition, a chunk also contains a {\em mapping} that indicates, for each record, which versions it belongs to (as a list of version-ids). Such a mapping is essential since a record may belong to multiple versions, and as discussed above, there is no easy way to identify which records belong to which versions. 

This design was motivated by the desire to address the shortcomings of the baseline approaches discussed above, by having several tuning knobs that could be used to adapt to different data and query workloads. The main reason behind chunking was to address the problem of too many queries. By keeping records that need to be retrieved together in a single chunk, we minimize the number of queries that need to be made to the backend store. At the same time, through appropriately setting the parameters, our system can easily emulate the behavior of the different baselines discussed above. For example, the ``sub-chunk'' approach can be easily emulated by forcing the partitioner to put all records with the same primary key in a single chunk, and keep different primary keys in separate chunks. However, in general, for mixed workloads, a hybrid solution ends up being ideal, where each chunk contains a few sub-chunks, each containing a subset of the records with the same primary key; such a partitioning not only reduces storage requirements by exploiting compression, but also reduces the number of queries that need to be made to the back-end.

\vspace{-8pt}
\begin{example}\label{ex:partition}
Table below shows two different partitionings for the data from Example~\ref{ex:composite_key}. To reconstruct version $V_1$ which contains 
$\langle K_0, V_0 \rangle,  \langle K_1, V_0 \rangle,  \langle K_2, V_0 \rangle,  \langle K_3, V_1 \rangle,  \langle K_4, V_1 \rangle$, we must retrieve
chunks $C_0, C_1, C_2, C_3$ for $\pp_0$, and chunks $C_0, C_1, C_2$ for $\pp_1$ (using the indexes discussed below). Overall, $\pp_1$ reduces the average number
of chunks to be retrieved per version by 0.6, and is thus a better option.

\begin{table}[h]
\centering
 \begin{tabular}{| c | c | c |}
  \hline
  Partition & $\pp_0$ & $\pp_1$ \\ \hline
  $C_0$ & $\{\langle K_0, V_0\rangle, \langle K_1, V_0\rangle\}$ & $\{\langle K_0, V_0\rangle, \langle K_1, V_0\rangle\}$ \\
  $C_1$ & $\{\langle K_2, V_0\rangle, \langle K_3, V_0\rangle\}$ & $\{\langle K_2, V_0\rangle, \langle K_3, V_0\rangle\}$ \\
  $C_2$ & $\{\langle K_3, V_1\rangle, \langle K_3, V_2\rangle\}$ & $\{\langle K_3, V_1\rangle, \langle K_4, V_1\rangle\}$ \\
  $C_3$ & $\{\langle K_4, V_1\rangle, \langle K_5, V_2\rangle\}$ & $\{\langle K_3, V_2\rangle, \langle K_5, V_2\rangle\}$ \\
  $C_4$ & $\{\langle K_3, V_4\rangle\}$ & $\{\langle K_3, V_4\rangle\}$ \\
  \hline
 \end{tabular}
\label{table:part}
\end{table}
\end{example}

\topic{Application Server (AS)}
The application server serves as the interface between the clients and the backend KVS, and comprises of three main modules described next. It 
uses the KVS for persisting any of its data structures. Multiple copies of AS could co-exist, with the standard caveat
that any data structures must be kept consistent across them (not currently supported in \ds).

AS currently provides a basic set of VCS commands. A user can {\em pull} any specific version by specifying its ID, or may {\em pull} the latest version in a branch (including the main {\em master} branch). Unlike a typical VCS, AS also provides the ability to retrieve partial versions or evolution history of a specific key as discussed in Section II(A). Any changes made by the user can be committed as a new version as discussed below.

\topic{Data Ingest Module}
Whenever a user commits a version, a {\em version-id} is generated by the system and is returned to the user after the commit process is complete. 
Even if two versions committed are exactly the same, the system will generate different version-ids for the two different commits (to account for different users, times at which they are committed, etc.). Due to the relatively large sizes of the datasets, the system requests only those records from the client that have changed, which in essence is the delta from the predecessor version. 
Thus the delta includes those records which have changed w.r.t. the previous version, records that are newly added and records that are deleted. If the client is unable to provide the delta, then the server needs to retrieve the prior version and perform a {\em diff} operation to check which records have been modified. However, in most settings, it is reasonable to assume that the client can do this.

Since updating the key-value store, and all the indexes, for every new version would be impractical, the received deltas are kept in a separate storage area, that are processed in a batch fashion by the data placement module. 
%

\topic{Data Placement Module}
This module is responsible for organizing the ingested data for efficient query processing. Once all the tuples ingested have been assigned a composite key, the data storage module scans through the records and places them into appropriate chunks using the underlying partitioning algorithm. In addition to placing the records, it is also responsible for constructing the version-record index for every chunk constructed and the version-chunk index that resides with the client for retrieving the versions. The chunks and associated indexes are stored in the KVS separately, in two distinct tables.

\begin{figure}[t]
\centering
\begin{tabular}{@{}cc@{}}
\subfloat[\label{subfig:p-v-c}] {\includegraphics[trim=5cm 10cm 5cm 10cm, width=2.2in]{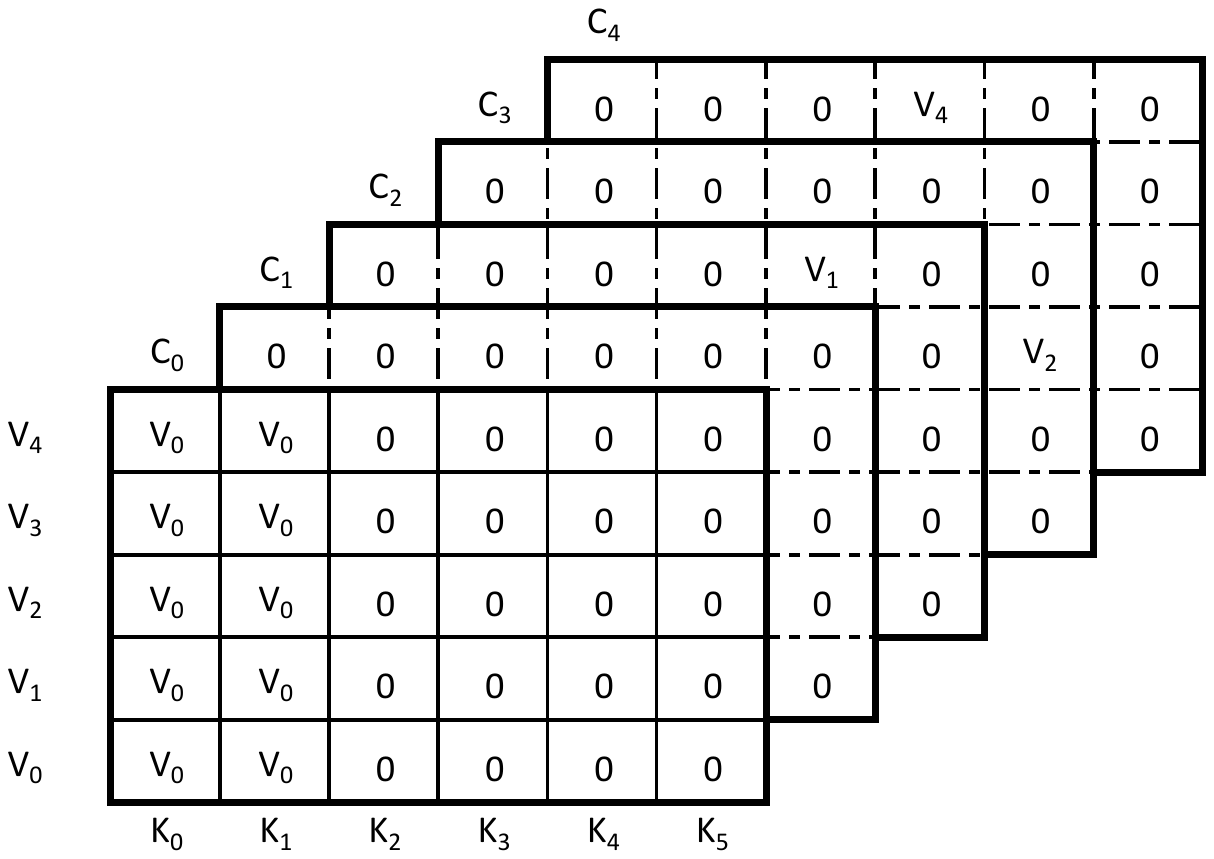}} & 
\subfloat[\label{subfig:p-c}] {\includegraphics[trim=6.5cm 9cm 3cm 10cm, width=1.9in]{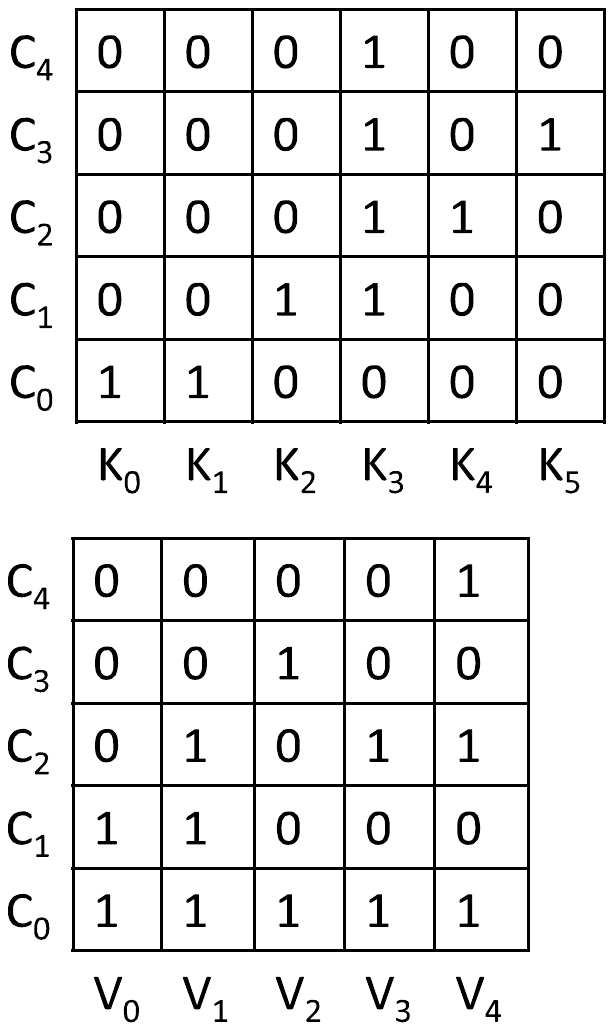}}
\end{tabular}
\vspace{-12pt}
\caption{(a) Entire 3D mapping; (b) Lossy projections maintained as indexes}
\label{fig:indexes}
\end{figure}

\topic{Indexes and Query Processing Module}
After the partitioning is completed, the system needs to know which chunks must be retrieved to extract the records belonging to a version. As discussed above, such an index is required even in the simplest approach, to be able to store any specific record only once even if it appears in multiple versions. Figure \ref{subfig:p-v-c} depicts the entire mapping, denoted $\mm_{|K| \times |V| \times |C|}$, between primary keys, version-ids, and chunks, that captures where each record is stored, and which versions contain it. 
The cells of this 3-dimensional matrix are annotated with version-ids that are required to construct the appropriate composite keys.
Specifically, the entry $\mm(K_i, V_j, C_k) = V_l$ if a {\em record} with composite key $\langle K_i, V_l\rangle$ is placed in {\em chunk} $C_k$ and belongs to {\em version} $V_j$; otherwise the entry is set to 0. This matrix is expected to be very large and highly sparse, but the information it depicts must be somehow maintained, either implicitly or explicitly, in the system.

We maintain this information as follows. First, with each chunk in the backend key-value store, we maintain the slice of the matrix corresponding to that chunk, $\mm^{C_i}$. This allows us to extract the records that belong to any specific version after the chunk has been retrieved from the backend key-value store. In aggregate, all of these ``chunk maps'' contain exactly the same information as $\mm_{|K| \times |V| \times |C|}$. Note that, the chunk maps will exploit the sparsity of the 2D matrix by using a {\em value list} representation of the matrix.

Second, in order to be able to decide what chunks to retrieve for a given query, we maintain two {\em lossy} projections of the matrix:
(1) a mapping between primary keys and chunks that tells us which chunks contain records for a given primary key, and (2) a mapping between versions and chunks that tells us which chunks contain records from a given version.  We use in-memory hashmaps to store these mappings. 

Query processing itself is straightforward given these indexes. We briefly summarize it below.

\stitle{Version Retrieval:} The second projection is consulted to identify which chunks need to be retrieved, and those chunks are retrieved by issuing queries in parallel to the backend store. After the chunks are retrieved, the chunk maps are used to extract the records that belong to that version.

\stitle{Record Evolution:} Similar to the above but the first projection is used instead.

\stitle{Range Retrieval/Record Retrieval:} Similar to ``index-ANDing'', both the projections are used here to obtain two lists of chunks, and all chunks in the intersection are retrieved from the backend store. Note that, it is possible for us to retrieve a chunk and, after analyzing the chunk map, discover that it contains no records of interest -- this is an artifact of these being lossy projections. 

The size of the {\em version-to-chunk mapping} is essentially the sum total version span across all versions, assuming the mappings are stored as adjacency lists. For dataset C0 in Table II (one of our bigger datasets), this results in a total index size of 11.25MB, compared to a total dataset size of 16GB after deduplicating. 
The size of the {\em primary key-to-chunk mapping} is governed by the number of primary keys and the number of different chunks they belong to, which in turn is depends on the size of the chunk and the degree of compression. The size of the map for dataset C0 ranges from 25MB to 75MB. Thus even with significantly larger datasets and numbers of versions, these indexes can easily fit in the large main memory machines that are available today. In fact, with larger datasets, we would typically use larger chunk sizes and sub-chunk sizes, both of which directly lead to lower index sizes. We further note that these sizes are before compressing the indexes themselves -- standard techniques from {\em inverted indexes literature} can be used to compress the adjacency lists without compromising performance.

\subsection{Formalizing the Optimization Problem}\label{ssec:formalism}

The key computational challenge here is deciding how to partition the records into chunks to minimize the storage cost and maximize the query performance (or minimize the {\em retrieval costs}). As we discussed in Section~\ref{ssec:tradeoff}, both the amount of data retrieved and the number of chunks retrieved are crucial performance factors from the perspective of querying, whereas compressing records by putting different records with the same primary key in the same chunk is crucial for minimizing storage costs. To achieve predictable performance, we made the following design decision. 

\stitle{(Fixed chunk size assumption).} All chunks are assumed to be approximately the same size, denoted $C$, with variations of upto 25\% allowed.

This variation in the chunk size gives us flexibility while assigning variable-sized records to chunks, and ensures that we are not forced to do frequent reorganization when adding new versions. We recommend that the specific percentage be chosen based on the ratio of the average record size and the chunk size, so that a small number of records could be added to an already full chunk while staying within the limit; for our datasets, 25\% ends up being a somewhat conservative number, and in our experimental evaluation, the chunks were rarely more than 5-10\% overfull.

This allows us to focus on the number of chunks retrieved for queries as the key performance metric. Formally, we define the {\bf span of a query} 
to be the number of chunks that must be retrieved to answer that query.

Let $n$ denote the total number of versions, $m$ denote the total number of distinct records in them, and $G$ denote the version graph depicting the relationships between the versions.  For a given partitioning (i.e., chunking), the {\em storage cost} and the {\em retrieval costs} can be calculated as follows.


\stitle{Storage Cost.} The total storage required is dominated by the chunks; the different indexes constitute a relatively small and largely fixed overhead. However, because of compression, it is hard to express the total storage required by the chunks analytically. Instead we use the {\em number of chunks required} as a proxy for the total storage cost. Since all chunks are about the same size, this faithfully captures the relative storage costs of different partitionings.

%

\stitle{Retrieval Costs.} For a query, let $\theta_i$ denote the total number of chunks that need to be accessed for answering it. The total retrieval cost is comprised of the {\em communication cost}, which in turn depends on the number of queries made to the backend ($\theta_i$) as well as the total number of bytes transferred, and the {\em CPU cost} of extracting the relevant records from the chunks. Once again, it is difficult to express this cost analytically; however, given the fixed chunk size assumption, the overall cost is largely proportional to $\theta_i$, and we use that as our retrieval cost metric.

Since there are two different objectives here, analogously to~\cite{BhattacherjeeCH15}, we can formalize optimization problems in different manners. However, our fixed chunk size assumption simplifies the problem somewhat if there is no compression.

\stitle{\underline{Case 1: No Record-Level Compression.}} In this case, the total number of chunks is approximately equal to the total number of bytes across all the records divided by the size of a chunk ($C$). Thus the optimization problem can simply be stated as minimizing the retreival cost for a query workload by appropriately assigning records to the chunks.

\stitle{\underline{Case 2: Record-Level Compression Allowed.}} In this case, the number of chunks required depends on how much compression can be obtained by grouping together the records with the same primary key. In this paper, we do not attempt to solve the problem in its full generality. Instead, we simplify the problem by assuming that a parameter, denoted $k$, is provided that controls how many records with the same primary key may be compressed together. ($k = 1$ corresponds to No Record-level Compression case). We use this parameter to partition the records with the same primary key into {\em sub-chunks} that are compressed together in a first phase. Then, the problem of assigning sub-chunks to chunks reduces to Case 1, since the total number of chunks required is once again fixed.

\stitle{Converting Version Graphs to Version Trees.} 
The following observation leads to the importance of version graphs in partitioning the records: A record (or a group of records) that appears in a version in a given branch of a tree can only be present in versions which are its descendants thereby allowing records present in different branches to be placed in different chunks, resulting in better partitioning decisions. In the next section, we discuss three different algorithms that partition the records into respective chunks. Except the shingles-based algorithm, the other algorithms use the version graph as a guide while creating the partitions. Those algorithms traverse the nodes of this graph in some particular order, read the records in the deltas and place them in the chunks.

\begin{figure}[t]
\includegraphics[trim=5cm 8cm 5.2cm 4.3cm, width=3in,clip]{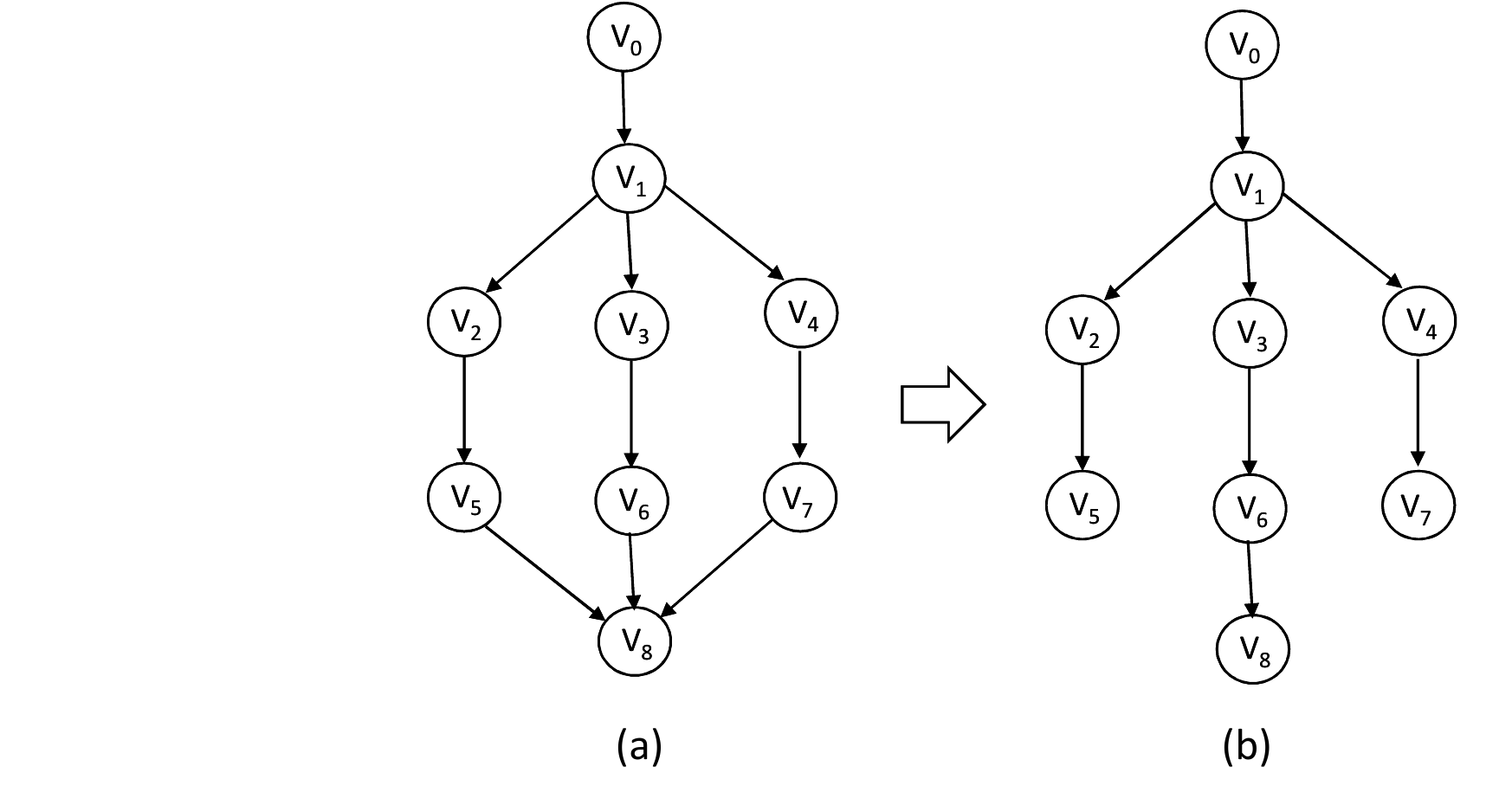}
\vspace{-16pt}
\caption{Converting a version DAG to a version tree}
\vspace{-16pt}
\label{fig:merge-vgraph}
\end{figure}

Due to the inherent complexity of the problem of partitioning, we use version graphs with no merges (henceforth referred to as {\em version trees}), in the subsequent partitioning techniques that use it. We use Figure~\ref{fig:merge-vgraph} to demonstrate the process of dealing with merges in version graph. Versions $V_5, V_6$ and $V_7$ form the list of parents of $V_8$. To convert the DAG to a tree, we choose a parent of $V_8$ arbitrarily (in this case $V_6$) retaining the edge between them while deleting the other two edges. In this process, there are records in $V_8$ that arrived exclusively from $V_5$ and $V_7$ which are renamed to make them appear as newly inserted records. 
\ul{This conversion is solely used during the partitioning phase and the original version graph is still available to any queries afterwards.}

\stitle{Connection to other problems}. The problem of partitioning records into chunks is closely related to the problem of identifying bicliques in a bipartite graph~\cite{FederM91}. In the current problem setting, the relationships between records and versions can be represented as a bipartite graph where there is an edge between a version and a record if that record appears in that version. We want to identify records that are present across a large number of versions from the bipartite graph. This is essentially finding the maximal biclique in the graph. In this case, we are interested in enumerating all maximal bicliques in the graph and then selecting bicliques that have disjoint set of records. However the problem of enumerating all maximal bicliques turns out to be NP-Hard~\cite{BuehrerC08}. Once we have the set of bicliques, we need to chunk them into bins of fixed size such that the number of bins required is minimized. This is the classical bin-packing problem and is NP-Hard.

The indexes used for recreating the versions have significant redundancy. Recall that an index for a version stores the keys of the records present in the version. In the current setting, the $\langle$chunk-id, record-key$\rangle$ pair can be used as a substitute of record keys. For two versions differing only by a few keys, the amount of redundancy is huge and therefore necessitates development of techniques for compressing the index. This problem is exactly equivalent to the problem of compression of posting lists~\cite{BlandfordB02, YanDS09}, where the version-id and the record keys correspond to the {\em term} and the {\em document-id's} in which the terms appear, respectively. This problem is also related to compressing graphs where the version-id and the record keys correspond to a graph node and its neighbors~\cite{AdlerM01, BoldiV04, ChierichettiKLMPR09}.

\subsection{Discussion}
In our discussion so far and in our prototype implementation, we assume that the backend KVS supports only a basic {\em get/put} interface. This raises the question of whether KV stores with richer functionality like {\em range} queries or {\em stored procedure} may negate the need for our approach. Although the trade-offs would be somewhat different, the key aspects of our approach are fundamental to the problem setting of maintaining versioned collections of records. Briefly, there are four key issues here: (1) exploiting overlap across versions, which we handle by not duplicating records, (2) retrieving a specific record from a specific version, which requires maintaining several large indexes efficiently, (3) too many queries problem, which we mitigate through careful assignment of records to chunks, and (4) compressing multiple versions of large records without compromising retrieval performance, which we handle through ``sub-chunking".  

As we discuss in Sections~\ref{ssec:tradeoff} and~\ref{ssec:toomanyq} , not addressing any of these will lead to substantial performance issues.
Hence, all four of the above proposed solutions must be present in any system that solves this problem effectively. In our prototype implementation that we presented in the paper, we assumed a simple key-value store (for maximum portability), and thus all four of those techniques had to be part of the RStore layer on top. Conceivably, a key-value store could handle one of those issues internally (i.e., implement one or more of our proposed techniques inside the key-value store), eliminating the need for them in RStore. \ul{However, we are not aware of prior work along these lines, and we consider the development of these approaches to be a lasting contribution of our work.}

Having support for range queries does not unfortunately remove the need for any of the four techniques as described above. The ``index" that tells us which records constitute a version still needs to be maintained in RStore; and the queries that will be posed against the backend key-value store will not be ``range" queries. For example, in the example in Figure 1, the list of records that constitute any of the versions cannot be captured as a range query on the composite key. Need for compression further complicates this because we need to retrieve ``sub-chunks" that contain the required records, and the sub-chunk IDs are effectively random.

Support for efficient large IN queries may help to some extent, depending on how they are implemented (in Cassandra, they are implemented by broadcasting to all data servers which leads to worse performance). In particular, that support will reduce the benefits of chunking, but not eliminate it. We still have the ``too many queries" problem, but {\bf internally to the key-value store}, i.e., \ul{there will be too many queries between the server that is collecting the query answer and the backend servers that host the data}. So a chunking approach may lead to improved performance. Unfortunately none of the key-value stores we investigated support large IN queries to investigate this properly.

Finally, ``stored procedures" cannot help here unless a large amount of the logic in RStore, including indexes, compression/decompression modules, and query module, is duplicated there. Even then, the ``too many queries" problem is still present between the query and the data servers.



\section{Partitioning Algorithms}\label{sec:exhaustive}


In this section, we present three different algorithms to solve the partitioning problem formalized in Section~\ref{ssec:formalism}. We begin with an adaptation of a standard {\em shingles}-based algorithm for finding bicliques, followed by two algorithms that exploit the inherent structure in the problem as embodied in the version graph. 

\subsection{Shingles-based partitioning}\label{ssec:shingle-based}

\resolved{Since we are using $k$ to denote the compression ratio, we should use $l$ or $m$ for the shingles.}
To minimize query spans, we want to place records together that are common to a large number of versions. This requires determining the similarity between records based on the versions they belong to. 
Here we adapt a standard technique for finding bi-cliques based on {\em shingles} or {\em min-hashing}, which provide an estimate of the similarity between large sets~\cite{BuehrerC08}. 
Briefly, for each set (here the set of versions that a specific record belongs to), we compute $l$ min-hashes, using hash functions $h_1, ..., h_l$; for each $h_i$, we apply the hash function to all the elements in the set and take the minimum of those as the $i^{th}$ min-hash. 
This gives us a list of $l$-shingles (min-hash values) for each record (Algorithm~\ref{alg:shingles-algo}). To compute the shingle ordering, we sort and order the records based on this list of shingle values in a lexicographical fashion. This ordering places records whose version sets have high similarity (i.e., overlap) in close proximity to each other. This shingle-based order is then used to place the records into the chunks (Algorithm~\ref{alg:part-shingle-algo}).

We also build the chunk maps, $\mm^{C_i}$ after all records have been assigned to their chunks. For every record in version $V_i$, we determine the chunk $C_i$ that it belongs to and add it to set of composite keys for that chunk. After scanning the full version, we 
visit every chunk that contained records from $V_i$ and write the version to composite key list to the corresponding chunk map file on disk. After this process is repeated for every version, we have the complete chunk map file for every chunk. The adjacency list in each chunk map file is then converted to a bitmap, compressed and stored in the KVS. Note that we use this algorithm for constructing the chunk maps for the subsequent partitioning algorithms as well.

\stitle{Complexity. } The complexity of the shingle-based technique for partitioning the records may be broken down as follows: 
\begin{denselist}
\item[1)] Constructing the record to version map takes $O(nm^{\prime})$ time which requires visiting every version and scanning every record in it, where $O(m^{\prime})$ is the average number of records in a version.
\item[2)] Next we compute the shingles for every record. If each record belongs to $O(n_{V})$ versions, then the time taken is $O(mn_{V})$. Note that the quantity $mn_{V}$ is $O(nm')$ as both of them denote the total number of records in the dataset.
\item[3)] Sorting the records based on $l$ shingle values takes $O(m\log m)$. Here the value of $l$ is a small constant.
\item[4)] Assigning the records to chunks can be done in $O(m)$ time.
\item[5)] Building the chunk maps takes $O(n(m^{\prime} + \rho_C))$ time. Here $\rho_C$ denotes the average number of chunks that records of any given version belongs to. Thus we have $\rho_C = O(m^{\prime})$ and the time complexity of constructing the chunks is $O(nm^{\prime})$.
\end{denselist}                                                                                  
Therefore the overall time complexity of this algorithm is $O(m\log m + nm^{\prime})$.
\resolved{There are too many symbols here -- what is $n_{V_s}$ (i.e., what does the subscript $s$ denote?). I think this can be simplified quite a bit.}


\begin{algorithm}[!t]
\DontPrintSemicolon
\SetKwInOut{input}{Input}\SetKwInOut{output}{Output}
{\small
\input{Set of versions $V \in S_i$, a family of $l$ pairwise-independent hash functions $H$}
\output{Shingle array of size $l$}
   $\mbox{shingles}[S_i] \leftarrow \{\}$\;
   \For {each $h \in H$} {
         $\mbox{shingles}[S] \leftarrow \{\mbox{shingles}[S], \mbox{min}_{v \in V}h(v)\}$\;
   }
   \Return $\mbox{shingles}$\;
}
\caption{Computing shingles for a set of versions}
\label{alg:shingles-algo}
\end{algorithm}

\begin{algorithm}[!t]
\DontPrintSemicolon
\SetKwInOut{input}{Input}\SetKwInOut{output}{Output}

{\small
\input{A set $r$ of records, version graph $G_t$, chunk capacity $C$}
\output{Set of chunks that partitions the records}
	\mbox{// Traverse the versions, scan records, construct record to version map}\;
        \mbox{// Compute Shingles for each record}\;
        \For {each $r_i \in r$} {
          $\omega_i = \mbox{ ComputeShingles}(r_i)$\;
        }
        \mbox{// Sort the records based on shingle values($\omega_i$)}\;
        \mbox{Sort($r$)}\;
        \For {each $r_i \in r$} {
          \mbox{// Assign records to chunks $C_j$ using the shingle-based sort-order}\;
          \If {$C_j < C$} {
            $C_j \leftarrow C_j \cup r_i$\;
          }
          \Else {
            \mbox{Create a new chunk and assign $r_i$}\;
	  }
	}
}
\caption{Shingle-based partitioning}
\label{alg:part-shingle-algo}
\end{algorithm}

\subsection{Bottom-Up Traversal}

\begin{figure}[t]
\includegraphics[trim=7cm 9.5cm 12cm 5cm, width=3in, clip]{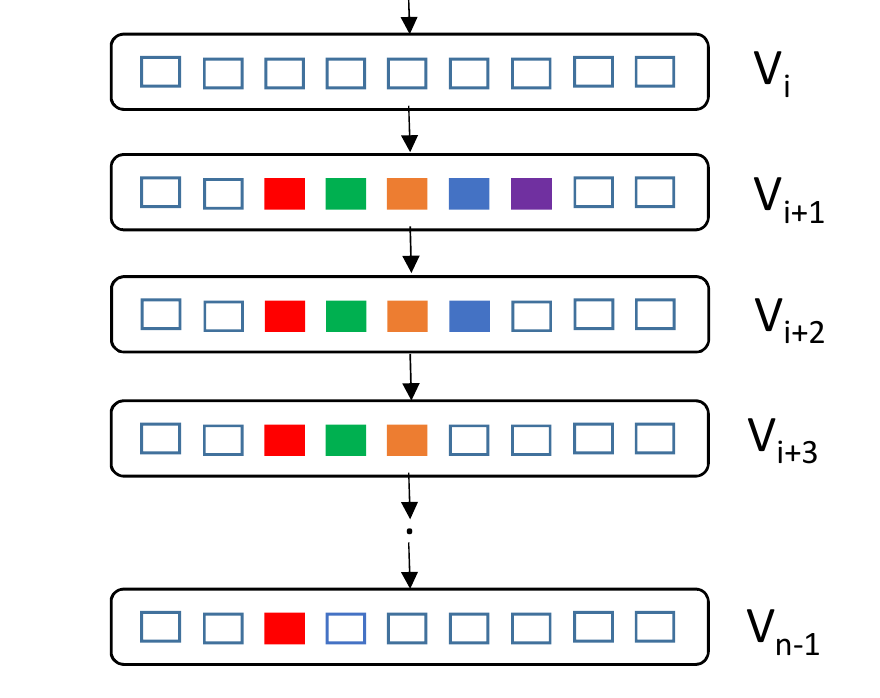}
\caption{Bottom-Up Partitioning for Linear Chains}
\label{fig:bottom-up-part}
\end{figure}

In this approach, we partition the records in the versions by traversing the version tree bottom-up\footnote{The Bottom-Up algorithm is inspired by~\cite{JansenKLS01} that gives an algorithm for partitioning a graph into two equal-sized partitions. In general, partitioning even trees is NP-hard~\cite{FeldmannF15}.}. The key idea here is to identify and chunk records that do not belong to versions above as we move up through the versions in the version tree. For simplicity, we will first describe the approach for 1-ary version trees and then extend it to general trees. Let us consider a version $V_i$ as depicted in Fig.~\ref{fig:bottom-up-part} which needs to be processed. Since we follow a bottom-up approach, the versions below $V_i$ in the version tree have already been processed. Let $S_i$ denote the set of records in $V_i$.  The collection of sets $\pi_{i+1} = \{S^{1}_{i+1}, S^{2}_{i+1}, \ldots, S^{p}_{i+1}\}$ contain the records that are returned by version $V_{i+1}$ and denote the following:
{\small
\setlength{\belowdisplayskip}{2pt} \setlength{\belowdisplayshortskip}{0pt}
\setlength{\abovedisplayskip}{2pt} \setlength{\abovedisplayshortskip}{0pt}
\begin{align*}
S^{1}_{i+1} &: \mbox{records present in $V_{i+1}$ but not in any version below.}\\
S^{2}_{i+1} &: \mbox{records present in $V_{i+1}, V_{i+2}$ but not in any version below.}\\
&\colon                    \\
S^{p}_{i+1} &: \mbox{records present in $V_{i+1}, V_{i+2}, \ldots, V_{i+p}$.}
\end{align*}
}%
Here $p$ denotes the number of versions from the current version (in this case $V_{i+1}$) up to the leaf version. Similarly, $V_i$ needs to return these sets to its parent $V_{i-1}$. In the present iteration, we compute the collection $\pi_i = \{S^{1}_{i}, S^{2}_{i}, \ldots, S^{p}_{i}\}$, where
{\small
\begin{align*}
S^{1}_{i} &: \mbox{records present in $V_{i}$ but not in any version below.}\\
&\colon                    \\
S^{p}_{i} &: \mbox{records present in $V_{i}, V_{i+1}, \ldots, V_{i+p}$.}
\end{align*}
}%
These sets are computed as follows:
{\small
\begin{align*}
S^{2}_{i} &= S^{1}_{i+1} \cap S_i, ~~~~~~~~ S^{3}_{i} = S^{2}_{i+1} \cap S_i\\
&\colon                   \\
S^{1}_{i} &= S_i \setminus (S^{2}_{i} \cup S^{3}_{i} \ldots \cup S^{p}_{i})
\end{align*}
}%
\noindent
It is also possible to express the sets in $\pi_i$ in terms of deltas. First, we will define deltas, discuss some of their algebraic properties and then describe the expressions. A delta $\Delta$ between two versions $V_i$ and $V_j$ is a set of records that may be split into two disjoint sets, a positive delta set, $\Delta^{+}$ and a negative delta set, $\Delta^{-}$. $\Delta_{ij}^{-}$ denotes the set of records that are present in $V_i$ but not in $V_j$, whereas $\Delta_{ij}^{+}$ denotes the set of records that are present in $V_j$ but not in $V_i$. It is easy to see that $\Delta_{ij}^{+} = \Delta_{ji}^{-}$ and $\Delta_{ij}^{-} = \Delta_{ji}^{+}$. For the following expression to hold, we require the deltas to be consistent~\cite{herac96}, i.e., $\Delta_{ij}^{+} \cap \Delta_{ij}^{-} = \phi$. The collection $\pi_i$ can expressed in terms of $\Delta$ as follows:
{\small
\begin{align*}
S^{1}_{i} &= \Delta_{i,i+1}^{-}, ~~~~~~~~~
S^{2}_{i} = \Delta_{i+1,i+2}^{-} \setminus \Delta_{i,i+1}\\
&\colon                    \\
S^{p}_{i} &: V_{n} \setminus \bigcup_{j=0}^{p-1}\Delta_{i+j,i+(j+1)}
\end{align*}
}%
Note that for the last term we have the whole version $V_n$ instead of a $\Delta^{-}$ since the last version does not have a $\Delta$ to some other version that captures the records that are exclusively present in version $V_n$. For general trees, computing $\pi_i$ changes slightly only for versions which have more than one child. In those cases $S^{1}_{i}$ is the union of the $\Delta^{-}$ between version $V_i$ and its children.

\noindent Given the collection of sets obtained from $V_{i+1}$ and the sets computed at $V_{i}$, it is now possible to determine the records that exclusively belong to certain versions, denoted by $\psi_i = \{\alpha^{1}_{i}, \alpha^{2}_{i}, \ldots, \alpha^{p}_{i}\}$. Thus we have,
{\small
\begin{align*}
&\alpha^{1}_{i} = S^{1}_{i+1} \setminus S^{2}_{i} \mbox{ (records present only in $V_{i+1}$)}\\
&\colon \\
&\alpha^{p}_{i} = S^{p}_{i+1} \setminus S^{p}_{i} \mbox{ (records present in $V_{i+1}, V_{i+2}, \ldots, V_{i+p}$)}
\end{align*}
}%
\begin{lemma}\label{lemma:lin-chain}
Given a linear chain of versions, we have $\bigcap_{j=1}^{p}\alpha^{j}_{i} = \phi$, at any version $i$.
\end{lemma}

\noindent Note that the records present in the sets from $\alpha^{1}_{i}$ to $\alpha^{p}_{i}$ are not present in any version from $V_i$ or above; so we can chunk these records. The records in set $\alpha^{p}_{i}$ must be chunked first, followed by those in $\alpha^{p-1}_{i}$ and so on. This is because records in $\alpha^{p}_{i}$ belong to $p$ consecutive versions, followed by records in $\alpha^{p-1}_{i}$ which belong to $p-1$ consecutive versions and so on, the chunking process at any given version starts filling a new chunk (or bin). This is to ensure that access to highly common records during version reconstruction is not split across multiple chunks, which in turn results in increasing the version span. The partial chunks that may get created at the end of every chunking step are merged at the end to reduce fragmentation. We demonstrate the chunking process through an example as follows.

\begin{example}
Consider Fig.~\ref{fig:bottom-up-part} where we have a linear chain of versions. Boxes represent records within versions and the colored boxes are the records which appear in $V_{i+1}$ and not in any prior version. Therefore the colored boxes represent the records in $\psi_i$ with the purple record representing $\alpha_{i}^{1}$, since they appear only in version $V_{i+1}$. Similarly, we have the blue record in $\alpha_{i}^2$ and so on. It is easy to see that the record in red must be chunked first, followed by the records in green and orange, then blue and finally purple. 
\end{example}

For general trees, the primary difference with the existing approach lies in processing versions with more than one child. Recall that at every version $V_i$, the child of $V_i$ returns $p$ different sets to its parent. If $V_i$ has $\lambda$ children, then it receives $\lambda\times p$ sets from its children. Unlike in linear chains (Lemma~\ref{lemma:lin-chain}), a given record may be present in more than one set (and no more than $\lambda$ sets, one from each child) for general trees. In the presence of multiple sets obtained from multiple children, ordering the records may be performed as follows:
\begin{denselist}
 \item[1)] For every record, assign a count based on the number of consecutive versions it belongs to. The count is added for records that appear in multiple sets.
 \item[2)] Sort the records.
\end{denselist}
A close approximation to the above technique may be obtained by considering sets of records that belong to similar number of consecutive versions. Therefore sets from different children that correspond to same number of consecutive versions, are chunked together. To deal with duplicate records, a hash-table is maintained to identify records that have already been chunked.


\subsubsection{Controlling the subtree of a version}\label{sssec: botup-subtree} 

The size of the subtree corresponding to a version in the tree dictates the amount of processing that needs to be done per version. For general trees, the size of subtrees is significantly larger compared to linear chains due to the presence of multiple branches per version on an average. In order to bound the amount of processing, we may choose to have at most $\beta$ nodes (or sets) in the subtree; the subtree can be reduced by merging nodes within it. Recall that each version in subtree corresponds to a set of records $S^{j}_{i+1}$ that $V_{i+1}$ returns to $V_i$. The merging involves the following steps:
\begin{denselist}
 \item[1)] Sort the leaves of the subtree by the sizes of the corresponding sets and store it in $L_{s}$.
 \item[2)] For every version $V_x$ in the sorted set:
 \begin{denselist}
 \item[a)] Merge the contents with its parent $V_p$. Remove $V_x$ from $L_{s}$. 
 \item[b)] If every child of $V_p$ have been merged, then include $V_p$ in $L_s$.
 \end{denselist}
 \item[3)] Repeat until the number of nodes in subtree is equal to $\beta$.
\end{denselist}
It is easy to see that a reduction in the size of the subtree reduces the total execution time of the \bt algorithm as the amount of processing per version is proportional to $\beta$. This may be true upto a certain $\beta$ as the overhead of merging the nodes may dominate for smaller values of $\beta$. However, with a decrease in a number of sets, the partitioning quality may also degrade, explained as follows. Consider that there are 10 sets below a version forming a linear chain and we want to determine  the records in $\psi_i$. We find that record $\langle K_i, V_i \rangle$ belong to  10 consecutive versions whereas record $\langle K_j, V_j \rangle$ belong to 5 consecutive versions, among other records. Therefore record $\langle K_i, V_i \rangle$ has a higher ordering than record $\langle K_j, V_j \rangle$ during the chunking process. Now, consider that $\beta=5$. In this case, both the records may be placed together; record $\langle K_j, V_j \rangle$ may be chunked with other records that have higher depth instead of $\langle K_i, V_i \rangle$, which leads to degradation of the partitioning strategy.

An outline of the bottom-up partitioning algorithm is provided in Algorithm~\ref{alg:bottom-up-partitioning}.

\stitle{Complexity. }  At every version, the number of set operations we perform is proportional to the the number of versions below it. Each set operation can be bounded by $O(m^{\prime})$ although in practice this is significantly less as this is proportional to the size of a delta. Thus the total complexity of set operations for all versions is $O(n\beta m^{\prime})$. The complexity of constructing the chunks and chunk maps is $O(nm^{\prime})$.

\begin{algorithm}[!t]
\DontPrintSemicolon
\SetKwInOut{input}{Input}\SetKwInOut{output}{Output}
{\small
\input{Version graph $G_t$, root version $V_r$ and deltas, sub-tree limit $\beta$, chunk capacity $C$}
\output{Set of chunks that partitions the records}
$\mbox{Bottom-Up}\;(V_r)$\;
\Return $\mbox{set of chunks}$\;
   $\mbox{Bottom-Up}\;(v)\;\{$\\
         \If {v is not null} {
           \For {each child $\in$ v} {
             $\mbox{Bottom-Up}\;(child)$\;
           }
           \mbox{// process the sub-tree $T_v$ rooted at $v$}\;
           \For {each version $V_j \in T_v$} {
             \mbox{compute $S^{j}_{v}$}\;
           }
           \mbox{// return set collection $\pi_{v}$ to parent of $v$}\;
           \mbox{// compute the records exclusive to $v$ and chunk them}\;
           \mbox{// adjust sub-tree $T_v$ if the size of sub-tree $> \beta$}\;
         }
   $\}$\\
}
\caption{Bottom-Up Traversal for Partitioning}
\label{alg:bottom-up-partitioning}
\end{algorithm}

\subsection{Depth-First/Breadth-First Traversal}\label{ssec:dfs}
To see if the benefits of the Bottom-up approach could be obtained using a simpler algorithm, we designed two algorithms which also 
use the version tree but make the partitioning choices greedily. 
These approaches traverse the version tree starting from the root in a depth-first or a breadth-first fashion, and chunk the records
as they are encountered. We illustrate this with an example.

%
%

\begin{figure}[t]
\includegraphics[trim=5cm 11cm 5cm 11cm, width=3in, clip]{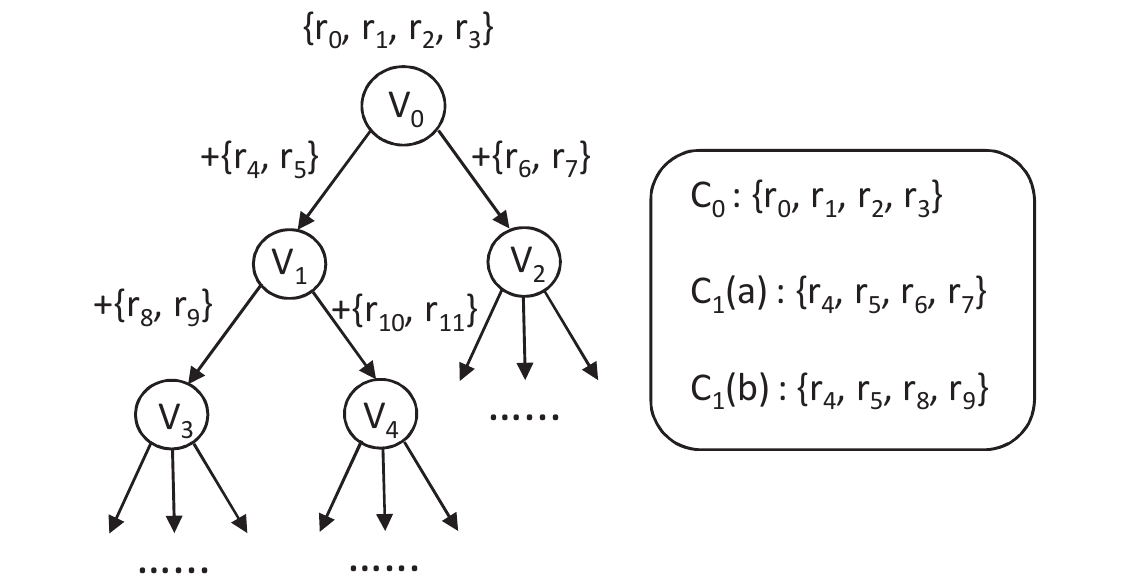}
\caption{Version Tree Partitioning (using DFS)}
\label{fig:vtree-part-dfs}
\end{figure}

\vspace{-6pt}
\begin{example}
Consider the version tree in Fig.~\ref{fig:vtree-part-dfs}, and assume the chunk size is 4 records. 
As the the root version $V_0$ is visited, all the records are placed in the first chunk $C_0$. Next, we visit one of the descendants of $V_0$, say $V_1$ and place the 2 records in the next available chunk $C_1$. Now, we have two options here, (a) visit version $V_2$ (breadth-first traversal) and place the two records in the remaining space in chunk $C_1$, (b) visit version $V_3$ (depth-first traversal) and place the two records in the remaining space in the chunk $C_1$. Note that going with option (a) implies that any descendant of $V_1$ will not access any of the records from $V_2$. Similarly, none of the descendants of $V_2$ will access any of the records added to chunk $C_1$(a) from $V_2$ resulting in the possibility of increasing the span of the versions. In contrast, option (b) admits all the descendants of $V_3$ to acces all the records in chunk $C_1$(b).
\end{example}

Assuming that most of the versions do not differ significantly from their parent version, traversing the version tree depth-first 
turns out to be more beneficial than breadth-first approach. An outline of the depth-first partitioning algorithm is provided in Algorithm~\ref{alg:dfs-partitioning}.

\stitle{Complexity. } The complexity of this algorithm is $O(nm^{\prime})$, where $O(nm^{\prime})$ is for traversing the all the records in each version. The complexity of chunk map construction is $O(nm^{\prime})$.

\begin{algorithm}[!t]
\DontPrintSemicolon
\SetKwInOut{input}{Input}\SetKwInOut{output}{Output}
{\small
\input{Version graph $G_t$, root version $V_r$ and deltas, chunk capacity $C$}
\output{Set of chunks that partitions the records}
   $\mbox{dfsStack}\leftarrow \{\}$\;
   \For {each $V_i \in G_t$} {
         $\mbox{visited}[V_i] \leftarrow \mbox{false}$\;
   }
   $\mbox{push dfsStack, }V_r$\;
   \While {dfsStack is not empty} {
         $u \leftarrow \mbox{peek dfsStack}$\;
           \If {u has child} {
             $v \leftarrow \mbox{getNextChild}(u)$\;
             \If {not visited[v]} {
               $\mbox{visited[v]} \leftarrow \mbox{true}$\;
               $\mbox{push dfsStack, v}$\;
               $\mbox{// read the delta and populate the chunk}$\;  
               \For {each record $r_i \in \Delta_{u,v}$} {
                 $C_i \leftarrow C_i \cup r_i$\;
                 $\mbox{// if } C_i \mbox{ is full then allocate a new chunk}$\;
               }
             }
           }
           \Else {
             $\mbox{pop dfsStack}$\;
           }
   }
   \Return $\mbox{set of chunks}$
}
\caption{Depth-First Traversal for Partitioning}
\label{alg:dfs-partitioning}
\end{algorithm}

\subsection{Partitioning Compressed Records}

\eat{
\begin{figure*}[t]
\centering
\includegraphics[trim=0cm 10.5cm 0cm 4cm, width=7in,clip]{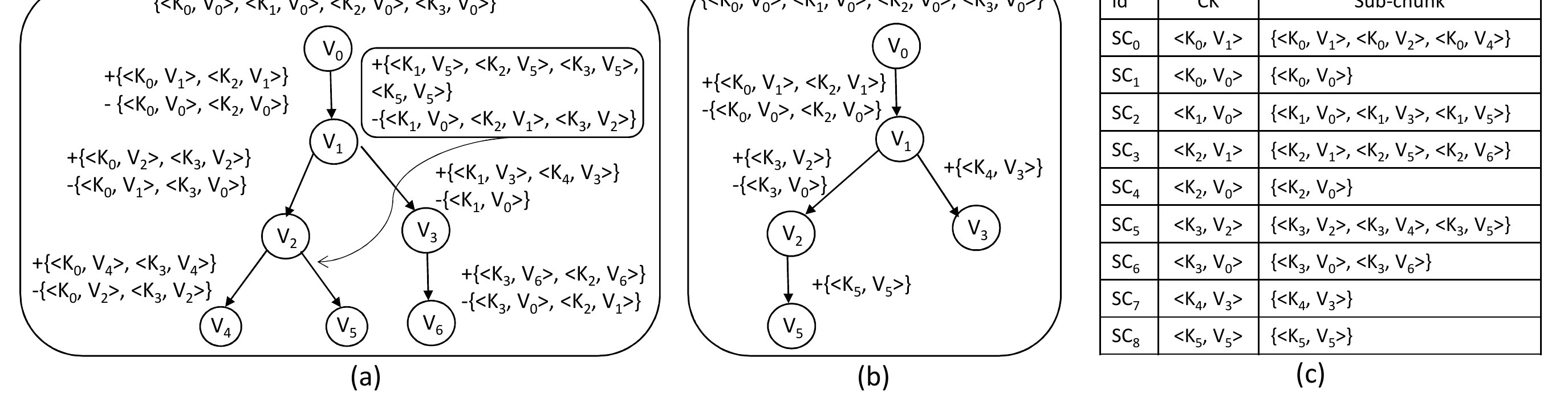}
\vspace{-12pt}
\caption{Partitioning compressed records: (a) Original Version Tree (b) Transformed Version Tree (c) Sub-chunk list with $\mu = 3$}
\vspace{-10pt}
\label{fig:compression-trans}
\end{figure*}
}

\begin{figure*}[t]
\vspace*{-20pt}
\centering
\includegraphics[trim=0.35cm 10.5cm 0.45cm 3cm, width=7in,clip]{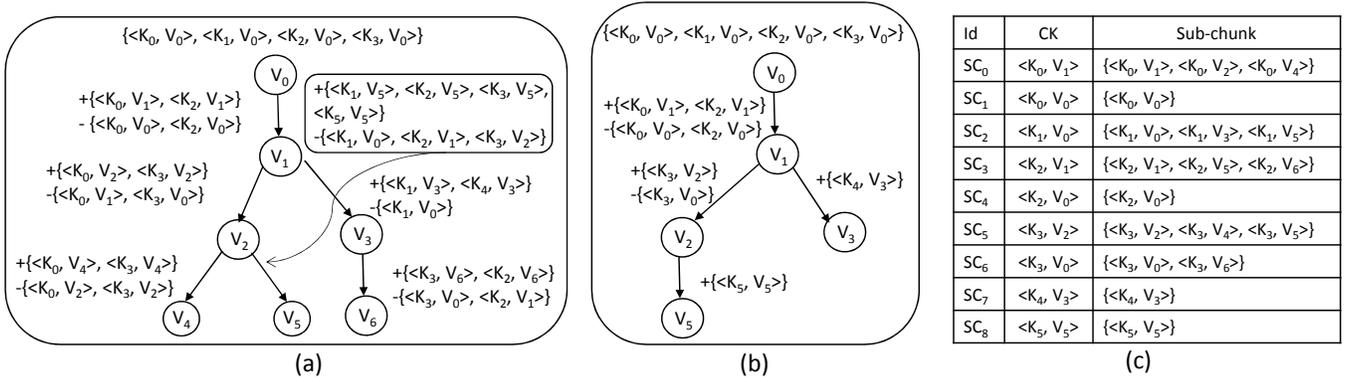}
\vspace{-12pt}
\caption{Partitioning compressed records: (a) Original Version Tree, (b) Transformed Version Tree, (c) Sub-chunk list with $k = 3$ -- CK indicates composite keys of the sub-chunks}
\vspace{-10pt}
\label{fig:compression-trans}
\end{figure*}

Next, we show how we handle the case where $k > 1$, i.e., we wish to exploit compression by putting together records with the same primary key in the same chunk.
As discussed in Section 2.5, we use a two-phase approach, where we first create the sub-chunks by grouping together records with the same primary key (with at most $k$ per sub-chunk), and then choose one of the partitioning algorithms discussed so far for the chunking itself by treating
the sub-chunks as records. Similar to records, we assign composite keys to these sub-chunks. One issue here is that, the original version tree may not be valid any more, and must be transformed (as discussed below) before the partitioning algorithms are invoked.

We impose the following constraint on any sub-chunk: the records that are grouped together are ``connected'' in the version tree, i.e., the versions that they belong to form a connected subgraph of the version tree. For example, in Figure \ref{fig:compression-trans}, we would
never group together $\langle K_1, V_3\rangle$ and $\langle K_1, V_5\rangle$, without $\langle K_1, V_0\rangle$ (their common ancestor). This is being done in order to boost compression as records are more likely to be similar to their parents than their siblings. Delta-encoding may be used to compress records within chunks; thus all the sibling records would be delta-ed against their common parent.

The sub-chunk creation algorithm proceeds by traversing the version tree bottom-up; at every version (excluding the leaf versions) we inspect its children and consider the records that originated in those child versions via inserts or updates. Every version can be assumed to have a collection of sets $\Psi$, each set in the collection $\Psi$ corresponds to a primary key that originated in that version. At any given version $V_i$, we either construct sub-chunks (for every primary key present in $V_i$ or any of its children) or delay the process until the next ancestor of $V_i$. Let $e(K_i)$ be a binary variable associated with a primary key $K_i$, which is 1 if $K_i$ is present in $V_i$, otherwise 0. Let $s(K_i)$ denote the count of the number of records for primary key $K_i$ across every child of $V_i$. If $e(K_i) = 1$ and $s(K_i) \leq k-2$, an union of the records is constructed and the set is added to $\Psi$ of $V_i$. However if $e(K_i) = 0$, instead of an union, the versions containing the records having primary key $K_i$ are added to the child list of the parent of $V_i$. In the situation, when $s(K_i) + e(K_i) \geq k$, we construct subchunks out of the largest set in $\Psi$ even though the set size may be less than $k$ and then recurse on the conditions mentioned above. 
We present an algorithm for sub-chunk construction at a given version $V_i$ (Algorithm~\ref{alg:sub-chunk-creation}). At every version $V_i$, we aggregate a list of primary keys that appears in $V_i$ or any of its children and denote it by $\sigma(V_i)$. For each $K_i$ in $\sigma(V_i)$, we execute the steps described earlier.

\begin{algorithm}[!t]
\DontPrintSemicolon
\SetKwInOut{input}{Input}\SetKwInOut{output}{Output}
{\small
\input{Version graph $G_t$, version $V_i$ and deltas, sub-chunk size $k$}
\output{Set of sub-chunks}
   
   \For {each $K_i \in \sigma(V_i)$} {
     \If {$e(K_i)\ = 1$} {
       \If {$s(K_i) = k-1$} {
         \mbox{construct sub-chunk.}\;
       }
       \ElseIf {$s(K_i) \leq k-2$} {
         \mbox{construct an union of records; add to $\Psi$}\;
       }
       \Else {
         \mbox{construct sub-chunk out of the largest set. Repeat.}\;
       }
     }
     \Else {
       \If {$s(K_i) \leq k-1$} {
         \mbox{add the children with $K_i$ to parent of $V_i$}\;
       }
       \Else {
         \mbox{construct sub-chunk out of the largest set. Repeat.}\;
       }
     }
   }
  
}
\caption{Sub-chunk Construction Algorithm at Version $V_i$}
\label{alg:sub-chunk-creation}
\end{algorithm}

\stitle{Transformed Version Tree.} The next step is to construct the transformed version tree $T_{VT}$ from the actual tree $O_{VT}$ by treating the sub-chunks as individual records. Each sub-chunk is assigned a representative composite key $\langle K_i, V_i\rangle$ which may lead to duplicate versions. 
Given the sub-chunks, the example below demonstrates the assignment of sub-chunks to records and the construction of the transformed version tree.
Different values of $k$ will lead to different transformations of $O_{VT}$ where each transformed version can be treated as a new dataset. 
The original partitioning algorithms can now be executed on these transformed datasets while taking into account the duplicate versions.

\begin{example}
Fig.~\ref{fig:compression-trans}(a) represents the original version tree and Fig.~\ref{fig:compression-trans}(b) represents the transformed version tree. The sub-chunks corresponding to $k = 3$ are extracted from $O_{VT}$ and are listed in Fig.~\ref{fig:compression-trans}(c) along with composite keys assigned to them. For deriving Fig.~\ref{fig:compression-trans}(b) from Fig.~\ref{fig:compression-trans}(a), we make a breadth-first traversal of $O_{VT}$ and at each version visit all the records that originated in that version. For every record, we pull up the corresponding sub-chunk that it belongs to and check whether it has already been used or not. For the root version in $V_0$, none of the sub-chunks corresponding to the records would have been assigned already. Therefore, the sub-chunks $SC_1, SC_2, SC_4$ and $SC_6$ are assigned the following representative composite keys: $\langle K_0, V_0\rangle$, $\langle K_1, V_0\rangle$, $\langle K_2, V_0\rangle$ and $\langle K_3, V_0\rangle$, respectively. Next, we move on to the records in $V_1$. We observe that $\langle K_0, V_1\rangle$ and $\langle K_2, V_1\rangle$ does not belong to the sub-chunks that were assigned composite keys in the previous step. So we assign $SC_0$ and $SC_3$ to $\langle K_0, V_1\rangle$ and $\langle K_2, V_1\rangle$, respectively. At $V_2$, we see that $\langle K_0, V_2\rangle$ is already in $SC_0$ whereas $\langle K_3, V_2\rangle$ isn't part of any sub-chunk that has been assigned already. Thus $\langle K_3, V_2\rangle$ is the representative composite key of $SC_5$. Similarly, $\langle K_4, V_3\rangle$ is assigned to $SC_7$. Next we visit $V_4$ and observe records that were new to it have already been a part of sub-chunk that have been assigned to its ancestors. In other words, $V_4$ has the same records as that of $V_2$ and hence $V_4$ is a duplicate of $V_2$ and hence $V_4$ is deleted. As we move on to $V_5$ we note that $\langle K_5, V_5\rangle$ has not assigned; thus $SC_8$ is assigned to $\langle K_5, V_5\rangle$. Finally, we observe that $V_6$ is a duplicate of $V_3$ and hence deleted. These steps result in the transformed version tree in Fig.~\ref{fig:compression-trans}(b).
\end{example}
\vspace{-2pt}

Creating the sub-chunks is expensive since the algorithm has to extract the sub-chunks by visiting all the different versions. For creating a single sub-chunk consisting of $k$ records, we have to visit $k$ different versions. To speed up this process, we first create the sub-chunks 
where we just have the composite keys of the records that form the sub-chunk. Thereafter, we concatenate the records from the versions and sort them by their primary keys on disk. Next, we scan the sorted record list and read all the records belonging a given primary key into memory. Since we maintain a record to sub-chunk map, we now create all the sub-chunks corresponding to the primary key, compress them and store them into a disk-based key-value store. Thus the sub-chunk creation is completed in a single pass over this sorted list of records.


\stitle{Complexity. }  The complexity of the sub-chunk construction algorithm
is $O(nm^{\prime} + m\log m)$, where $O(nm^{\prime})$ is for traversing the all
the records in each version and the second component is for sorting the unique
records for sub-chunk extraction. The complexity of chunk map construction is
$O(nm^{\prime})$.

\section{Online Partitioning}
The main challenge with keeping the partitioning up-to-date with every new version is that, even if a version $V_c$ differs from its parent version $V_p$ by just a few records, all the chunks that contain $V_p$'s records need to be updated (if only to update the chunk maps). As discussed earlier, we instead incorporate new versions in a batched fashion, by maintaining the deltas corresponding to the new versions in a separate write store, called a {\em delta store}, and by using an adapted version of a partitioning algorithm when the number of versions reaches a certain size (called the {\bf batch size}, a user-configurable parameter).

To exploit the possibly high overlap across versions in the current batch, we compute a union of the chunk maps that need to be updated and then update every chunk map only once per batch. In order for a chunk map to be updated if it already exists, it has to be fetched from the KVS, updated and then written back again. Instead, every time a chunk map needs to be updated per batch, we recreate the chunk index from scratch and then write it back to KVS, saving the cost of fetching the chunk indexes from the KVS. This is possible by maintaining the required indexes around due to its small memory footprint. The complexity of the background process is determined by the size of the batch and the choice of the partitioning algorithm. In general, a smaller batch size would result in faster partitioning, however the quality of partitioning degrades with respect to a larger batch as more versions in a batch is beneficial for making better record placement decisions. Note that we do not re-partition records once they have been partitioned, however record re-partitioning, although expensive, may result in improving the overall version span. We leave this problem for future work.

\section{Experiments}

In this section, we present a comprehensive evaluation of the \ds system. 
We use a distributed installation of Apache Cassandra across upto 16 nodes for storing the partitioned records and their associated indexes. 
Each node has 16 GB of main-memory. We ran our experiments on a 2.2 GHz Intel Xeon CPU E5-2430 server with 64GB of memory, running 64-bit RedHat Enterprise Linux 6.5.

\subsection{Datasets} 
We use a collection of synthetically generated datasets for the experiments. For each dataset, we first generate a corresponding version graph by starting with a single version, and then generating a set of modifications to it using the method outlined in~\cite{BhattacherjeeCH15}, which closely follows real-life version graphs.
Thereafter, we create a set of records for the base (root) version where each record is created as a JSON document. Every record in the base version is assigned an auto-incremented primary key and a randomly generated value of the requisite size. Each of the other versions is generated by updating or deleting a set of records in its parent, or inserting new records. 
The selection of records for updating and deleting either follows a random or a skewed (Zipf) distribution. 

We have generated a wide spectrum of version graphs and corresponding datasets that mimics real-world use cases. They differ primarily along five factors: 1) {\em branching factor} (linear to highly branched), 2) {\em average version graph depth} (56 to 300),  
3) {\em nature and percentage of updates} (random vs skewed updates with $1-50\%$ change), 4) {\em number of records in a version} (from a few thousand to hundreds of thousands of records), and 5) {\em number of versions} (from a few hundred to several thousand). The size of the records in the dataset also vary widely from a few bytes to several kilobytes. The number of unique records in the dataset varies from a little more than 1M records to around 17M records and total size of a dataset varies from $\approx30$ GB to close to 1 TB. We refer to Table~\ref{table:data-desc} for a detailed description of the datasets.

\begin{table*}[!ht]
\small
\centering
 \begin{tabular}{| c | c | c | c | c | c | c | c | c |}
  \hline
  Dataset & \#versions &  Avg. depth & $\sim$\#records/version & \%update & Update Type & \#unique records & Size of unique records (in GB)  & Total size (in GB) \\ \hline
  A0 & 300 & 300 & 100K & 50 & Random & 12355366 & 11.9 & 31.67 \\ \hline
  A1 & 300 & 300 & 100K & 5 & Skewed & 1510097 & 5.77 & 140.14 \\ \hline
  A2 & 300 & 300 & 100K & 5 & Random & 1343434 & 5.14 & 141.26 \\ \hline
  B0 & 1001 & 293.5 & 100K & 5 & Skewed & 4175023 & 8 & 192.24\\ \hline
  B1 & 1001 & 293.5 & 100K & 5 & Random & 4216366 & 8.07 & 193.77\\ \hline
  B2 & 1001 & 293.5 & 100K & 10 & Random & 8349864 & 8.02 & 195.69\\ \hline
  C0 & 10001 & 143 & 20K & 10 & Random & 16532342 & 15.95 & 196.46\\ \hline
  C1 & 10001 & 143 & 20K & 1 & Random & 1758517 & 1.69 & 193.01\\ \hline
  C2 & 10001 & 143 & 20K & 5 & Skewed & 8169026 & 7.87 & 193.05\\ \hline
  D0 & 10002 & 94.4  & 20K & 10 & Random & 16621314 & 16.03 & 196.48\\ \hline
  D1 & 10002 & 94.4  & 20K & 1 & Random & 1773281 & 1.71 & 193.07\\ \hline
  D2 & 10002 & 94.4  & 20K & 5 & Skewed & 8195193 & 7.90 & 193.09\\ \hline
  E & 10001 & 170 & 20K & 10 & Random & 16524584 & 78.96 & 972.84\\ \hline
  F & 1001  & 56  & 100K & 20 & Random & 16665072 & 79.64 & 981.11\\
  \hline
 \end{tabular}
 \caption{Description of the datasets used in experiments}
\label{table:data-desc}
\end{table*}


\subsection{Evaluation of Partitioning Algorithms}

\begin{figure*}[t]
\vspace*{-40pt}
\centering
\begin{tabular}{@{}cc@{}}
\subfloat[\label{fig:tot-vspan}]{\includegraphics[trim=4cm 7.5cm 2cm 5cm, width=4.5in]{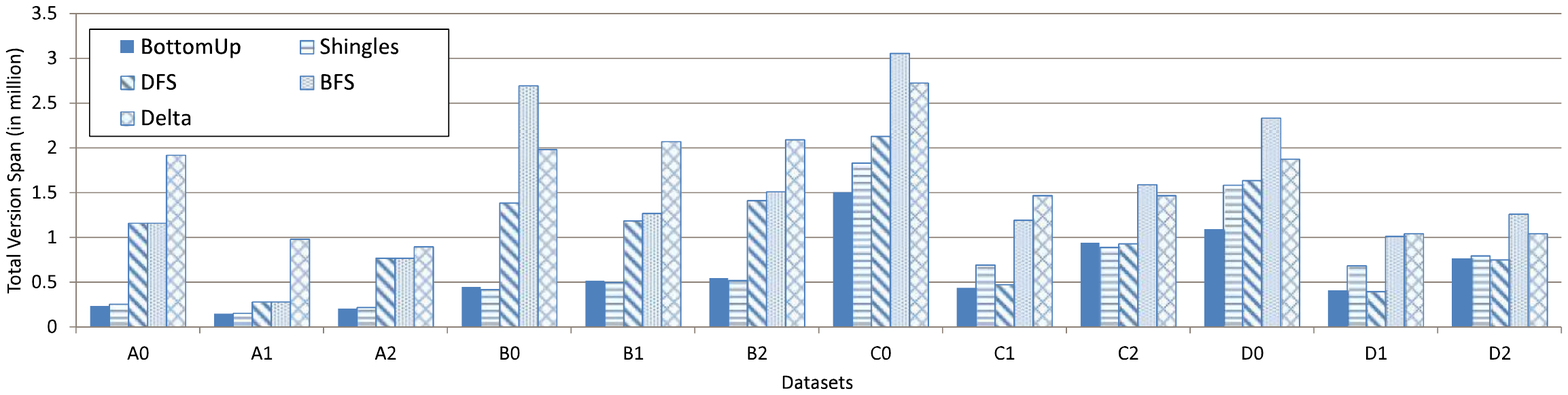}} &
\subfloat[\label{fig:tot-vspan-large}]{\includegraphics[trim=3cm 4cm 8cm 1cm, width=1.6in]{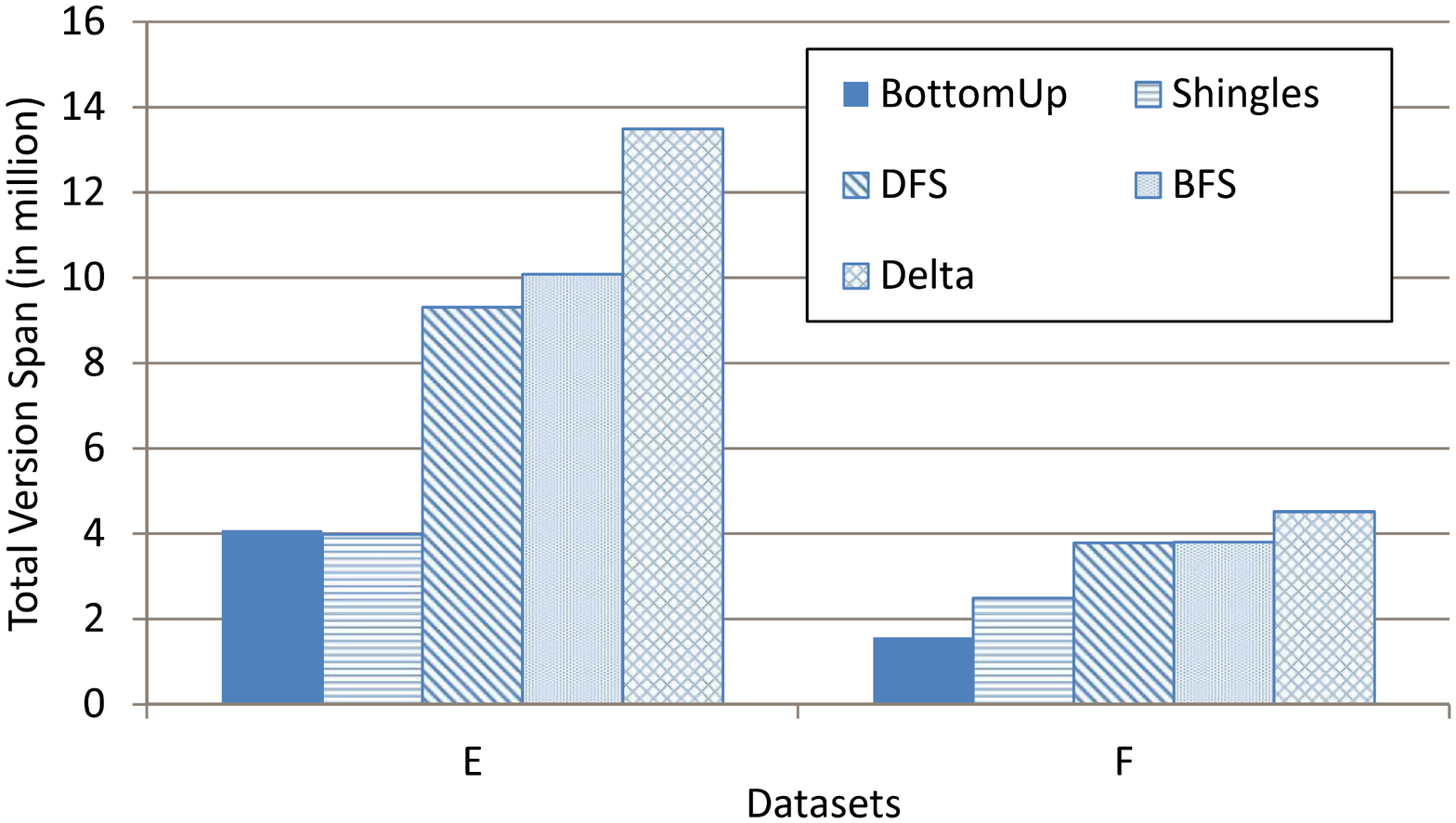}} 
\end{tabular}
\caption[]{Comparison of Total Version Span (without compression) }
\vspace{-10pt}
\label{fig:tot-vspan-all}
\end{figure*}

\stitle{Comparison based on Total Version Span}. We begin with comparing the performance of the partitioning algorithms: \bt, \sh, \dfs, and \bfs. 
Here, we use the total version span (i.e., the total number of chunks retrieved for reconstructing all versions) for comparing the algorithms while fixing the chunk size to 1MB (we chose this chunk size since it provides a good balance between the number of queries and amount of data retrieved). In addition to algorithms that partition the record space for minimizing the version span, we also show performance of the \deltat baseline. We omit the \subc baseline since the total version span for that
approach is very high (all chunks must be retrieved for any version query). 

In Fig.~\ref{fig:tot-vspan-all}, we observe that \bt, \sh and \dfs outperform \deltat across all datasets, thus establishing that \deltat is inferior as a technique for handling keyed datasets (\bt outperforms \deltat by upto $8.21\times$ and on an average by about $3.56\times$ across all datasets). The performance of \sh degrades with a decrease in the average depth of the version trees, while that of \dfs improves. {\bf However unlike \bt, none of these techniques perform uniformly well across all datasets}. \bfs is always worse than \dfs (for reasons described in Section~\ref{ssec:dfs}) except for linear chains when they reduce to the same technique.

\begin{figure}[t]
\vspace{-25pt}
\mbox{\ }
\hspace{.5in}
\includegraphics[trim=6cm 4.5cm 8cm 1cm, width=2in]{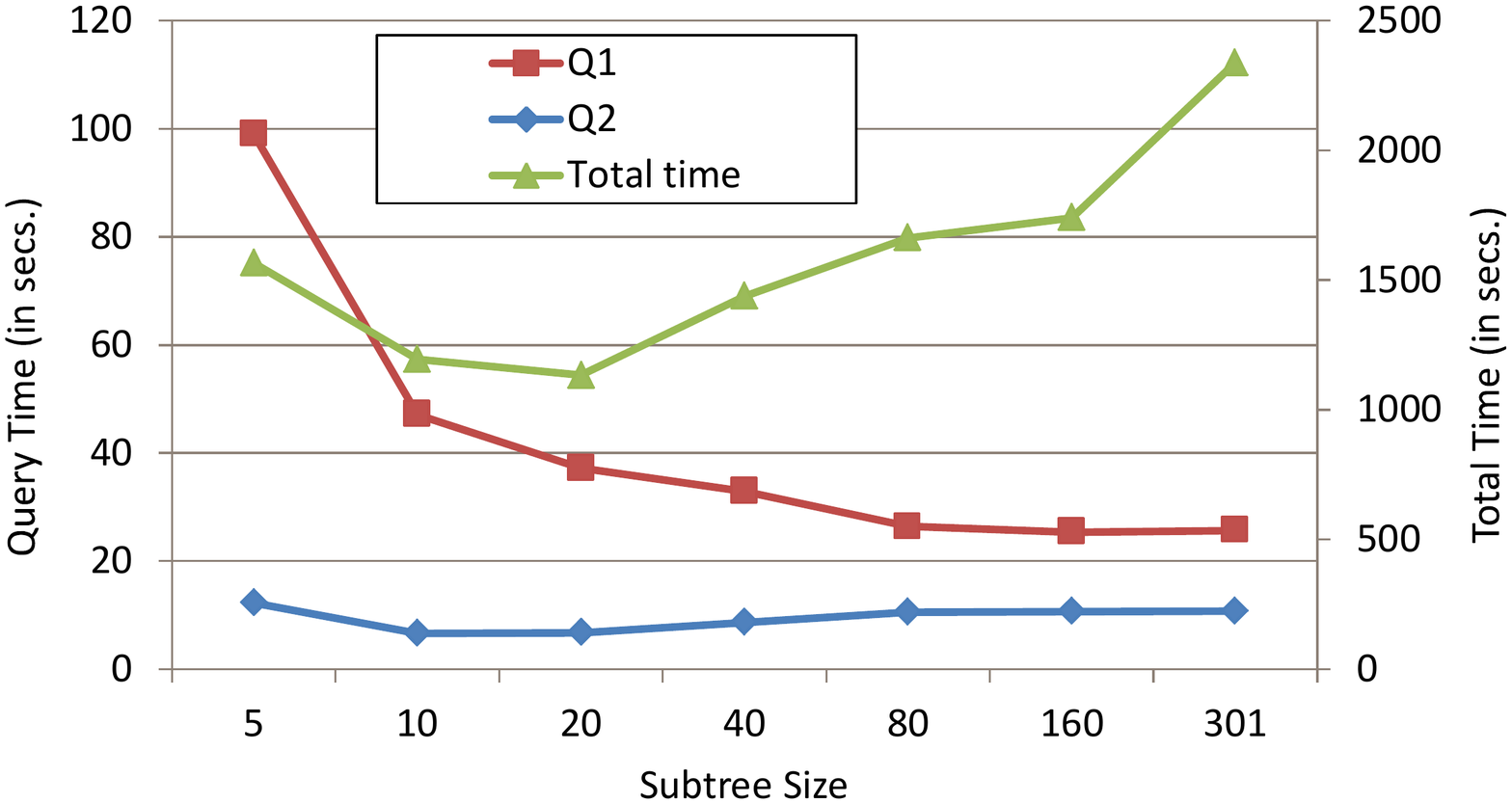}
\caption{Effect of sub-tree size on performance of \bt (Dataset B0)}
\vspace{-15pt}
\label{fig:subtree-G}
\end{figure}

\stitle{Effect of Subtree size on performance}. We vary the size of the subtree ($\beta$) \bt and observe the total version span (Fig.~\ref{fig:subtree-G}). As the size of the subtree decreases, the total version span increases as explained in Section~\ref{sssec: botup-subtree}. The total time taken by the algorithm first decreases with decrease in subtree size (due to decrease in processing per node) and then increases. The increase in total time for $\beta < 20$ in Fig.~\ref{fig:subtree-G} can be attributed to increased processing time for merging the nodes. As $\beta$ decreases the number of nodes needed to merge also increases.

\subsection{Effect of Compression on Partitioning}

\begin{figure*}[t]
\centering
\begin{tabular}{@{}ccc@{}}
\subfloat[\small Dataset A0, $P_d = 10\%$\label{subfig:a10}]{\includegraphics[trim=3cm 9.5cm 2cm 9cm, width=2.3in]{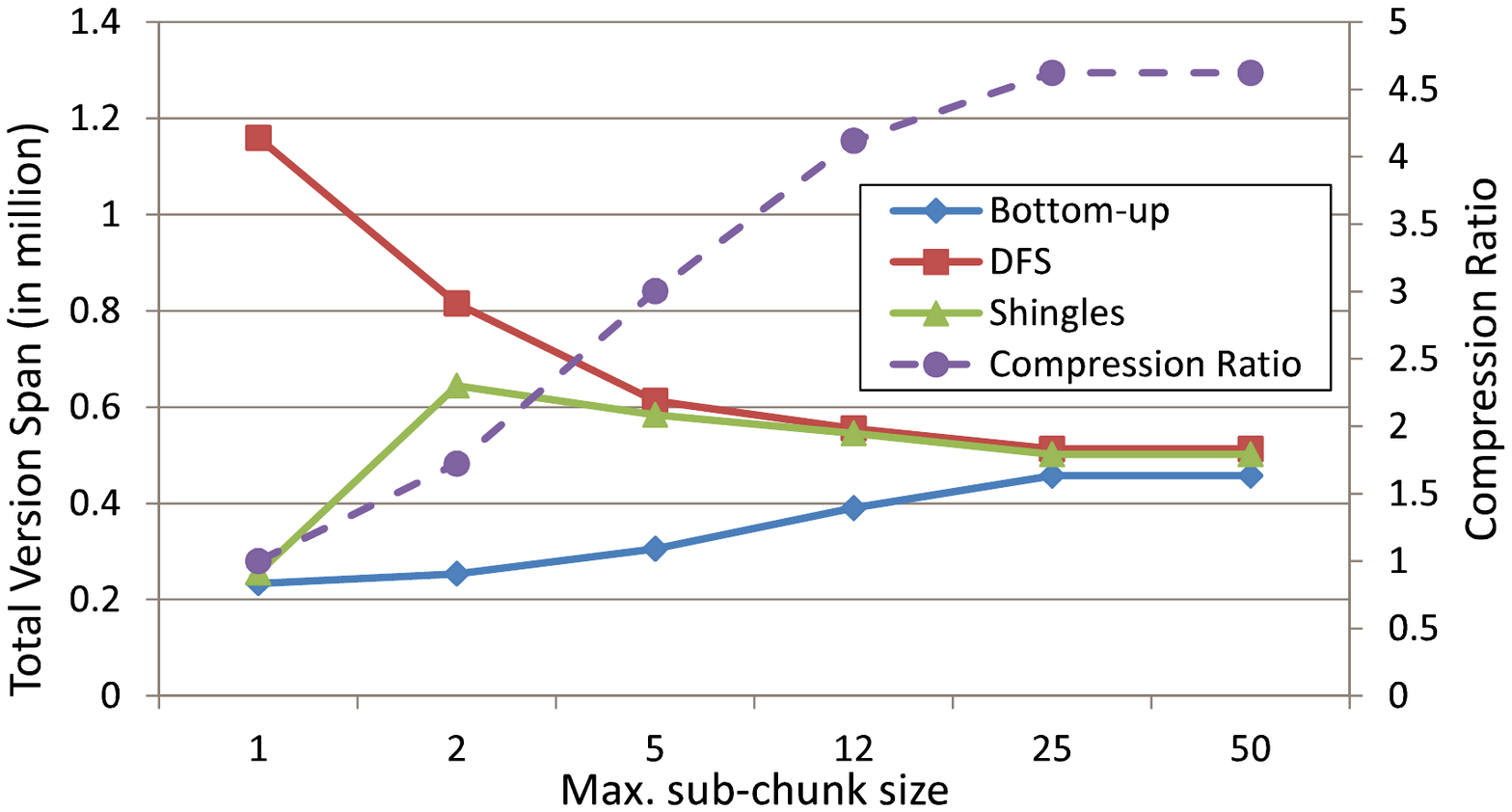}} & 
\subfloat[\small Dataset A0, $P_d = 5\%$\label{subfig:a5}]{\includegraphics[trim=3cm 9.5cm 2cm 9cm, width=2.3in]{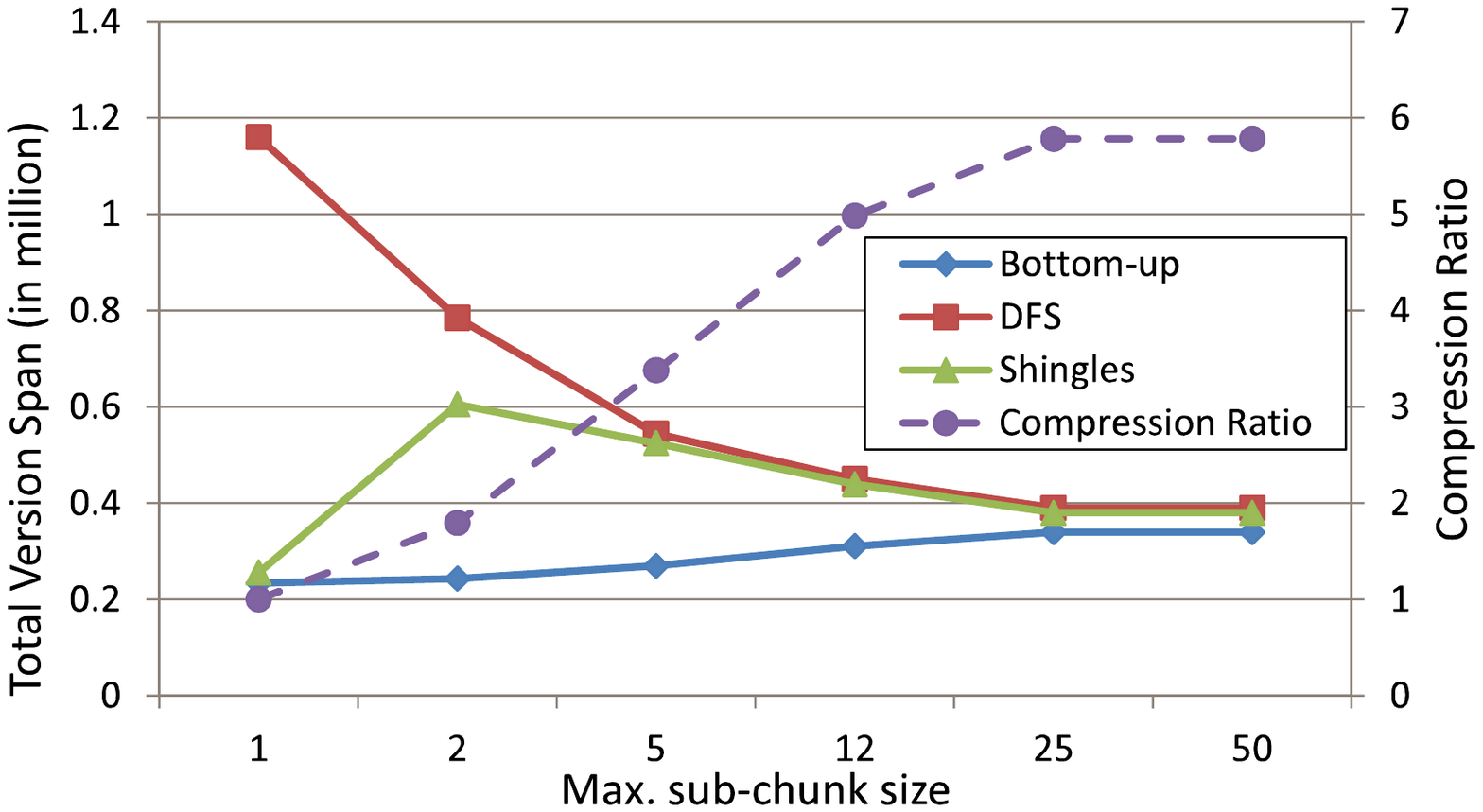}} &
\subfloat[\small Dataset A0, $P_d = 1\%$\label{subfig:a1}]{\includegraphics[trim=3cm 9.5cm 2cm 9cm, width=2.3in]{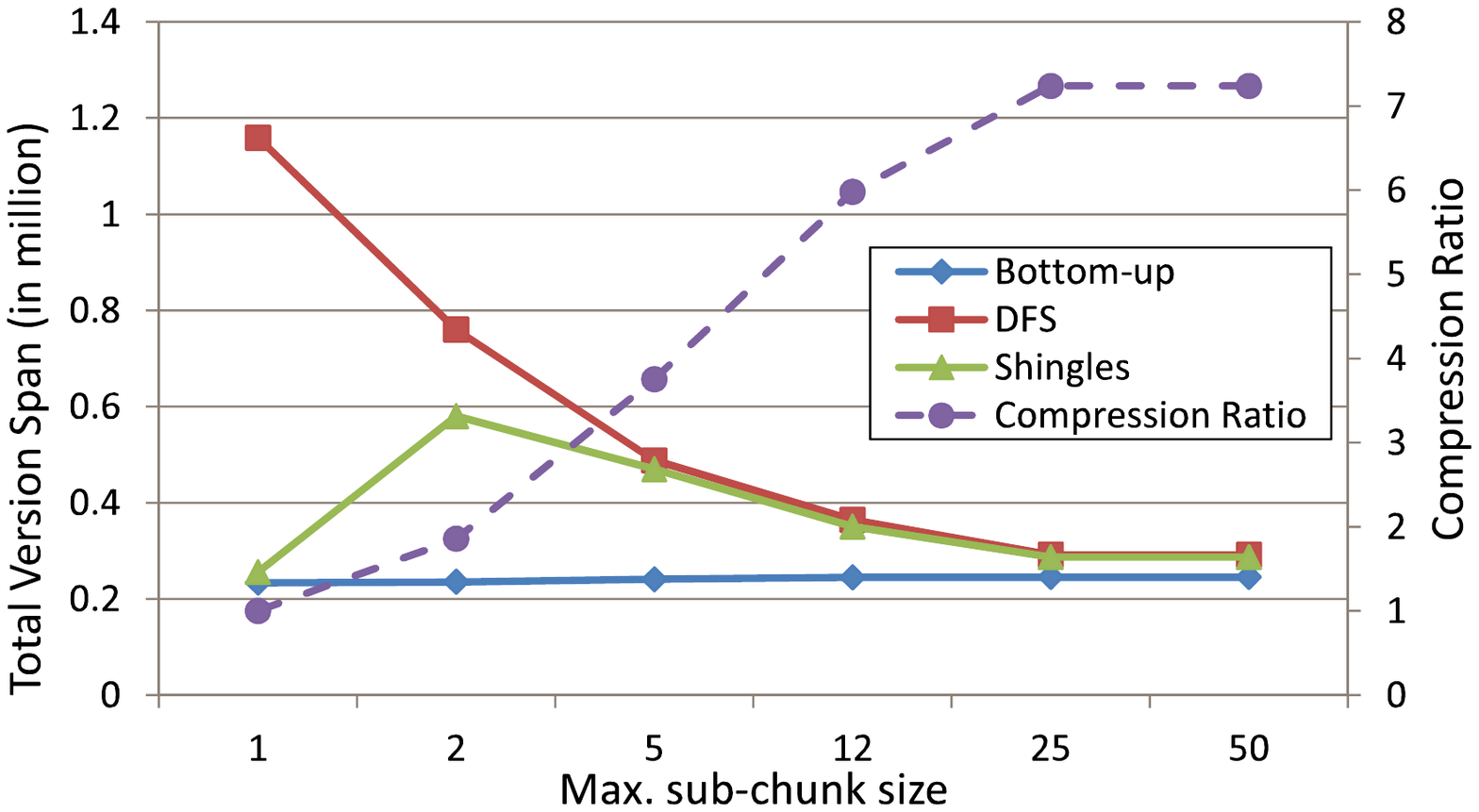}} \\[-2pt]
\subfloat[\small Dataset C0, $P_d = 10\%$\label{subfig:c010}]{\includegraphics[trim=6cm 6.5cm 5cm 10cm, width=2.3in]{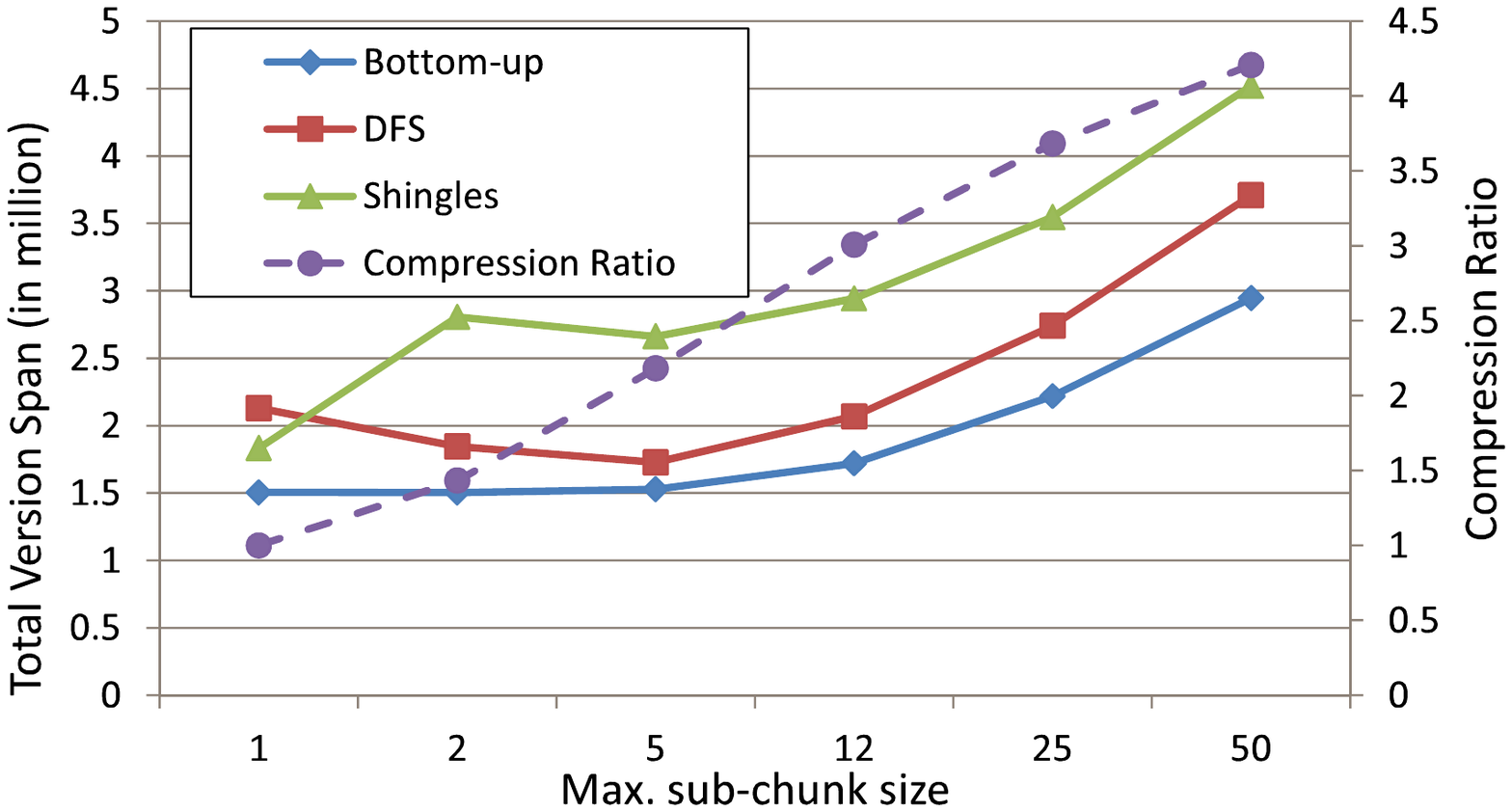}} &
\subfloat[\small Dataset C0, $P_d = 5\%$\label{subfig:c05}]{\includegraphics[trim=3cm 9.5cm 2cm 10cm, width=2.3in]{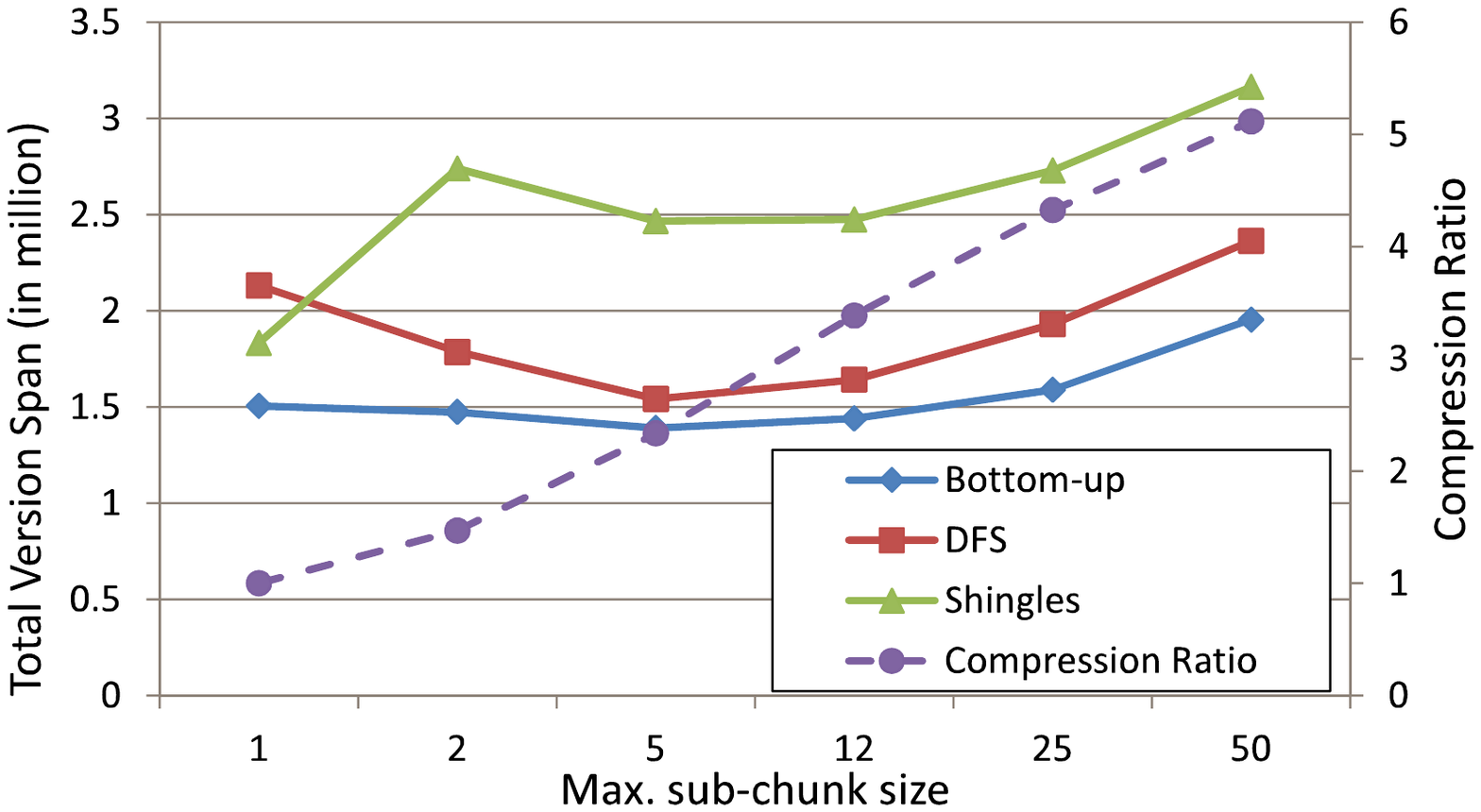}} &
\subfloat[\small Dataset C0, $P_d = 1\%$\label{subfig:c01}]{\includegraphics[trim=3cm 9.5cm 2cm 10cm, width=2.3in]{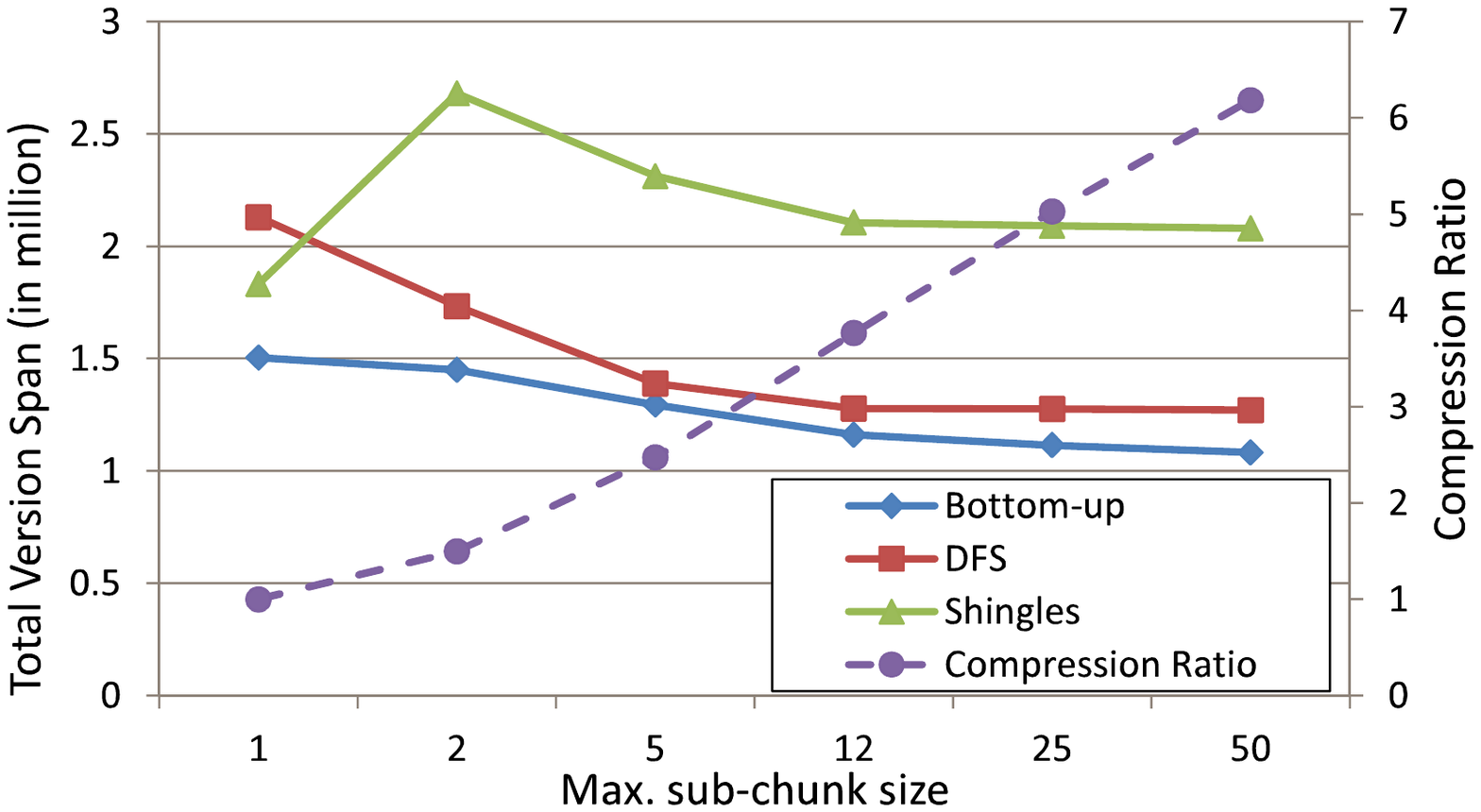}}\\[-6pt]
\subfloat[\small Dataset D0, $P_d = 10\%$\label{subfig:d010}]{\includegraphics[trim=6cm 6.5cm 5cm 6cm, width=2.3in]{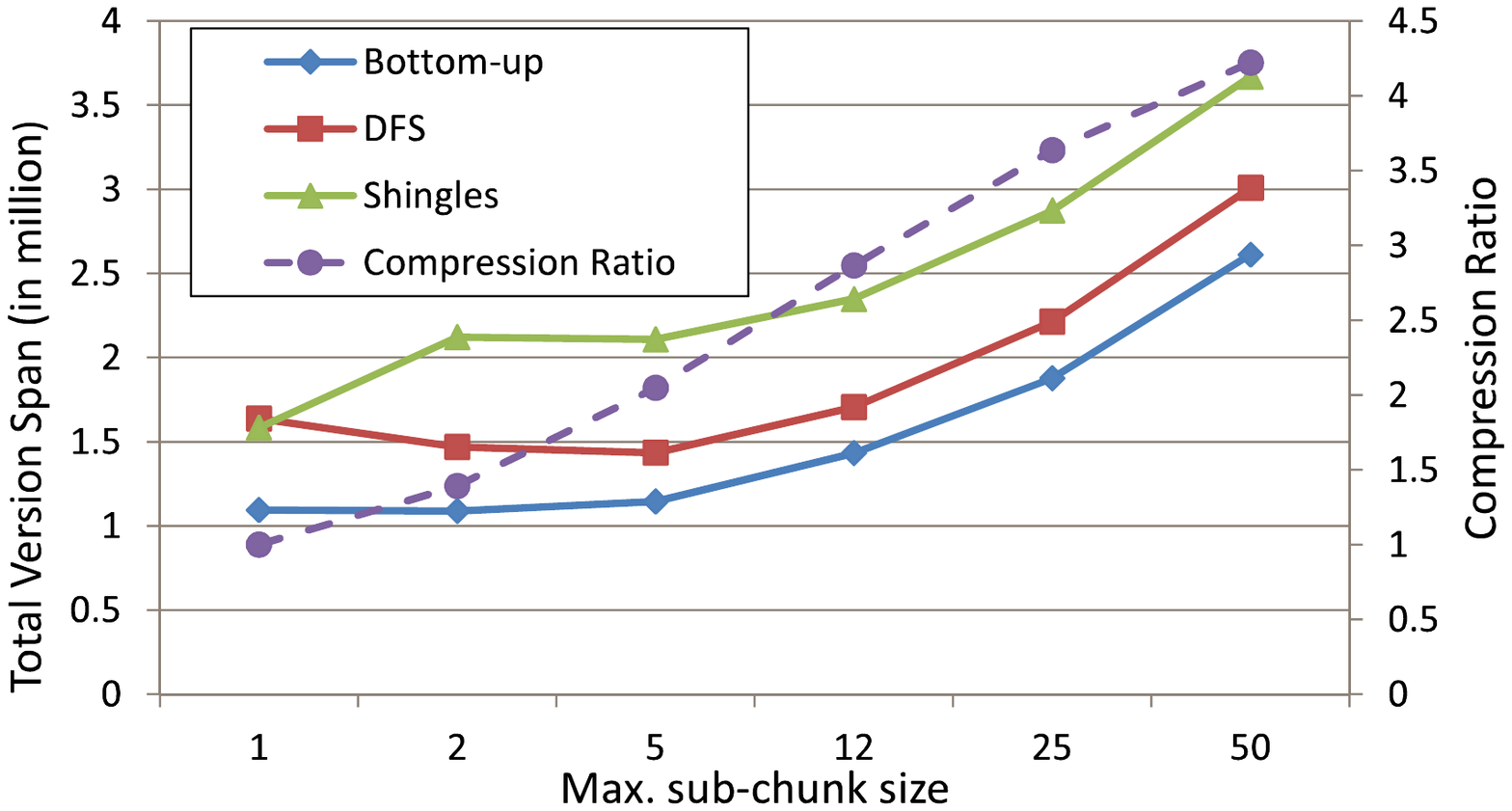}} &
\subfloat[\small Dataset D0, $P_d = 5\%$\label{subfig:d05}]{\includegraphics[trim=6cm 6.5cm 5cm 6cm, width=2.3in]{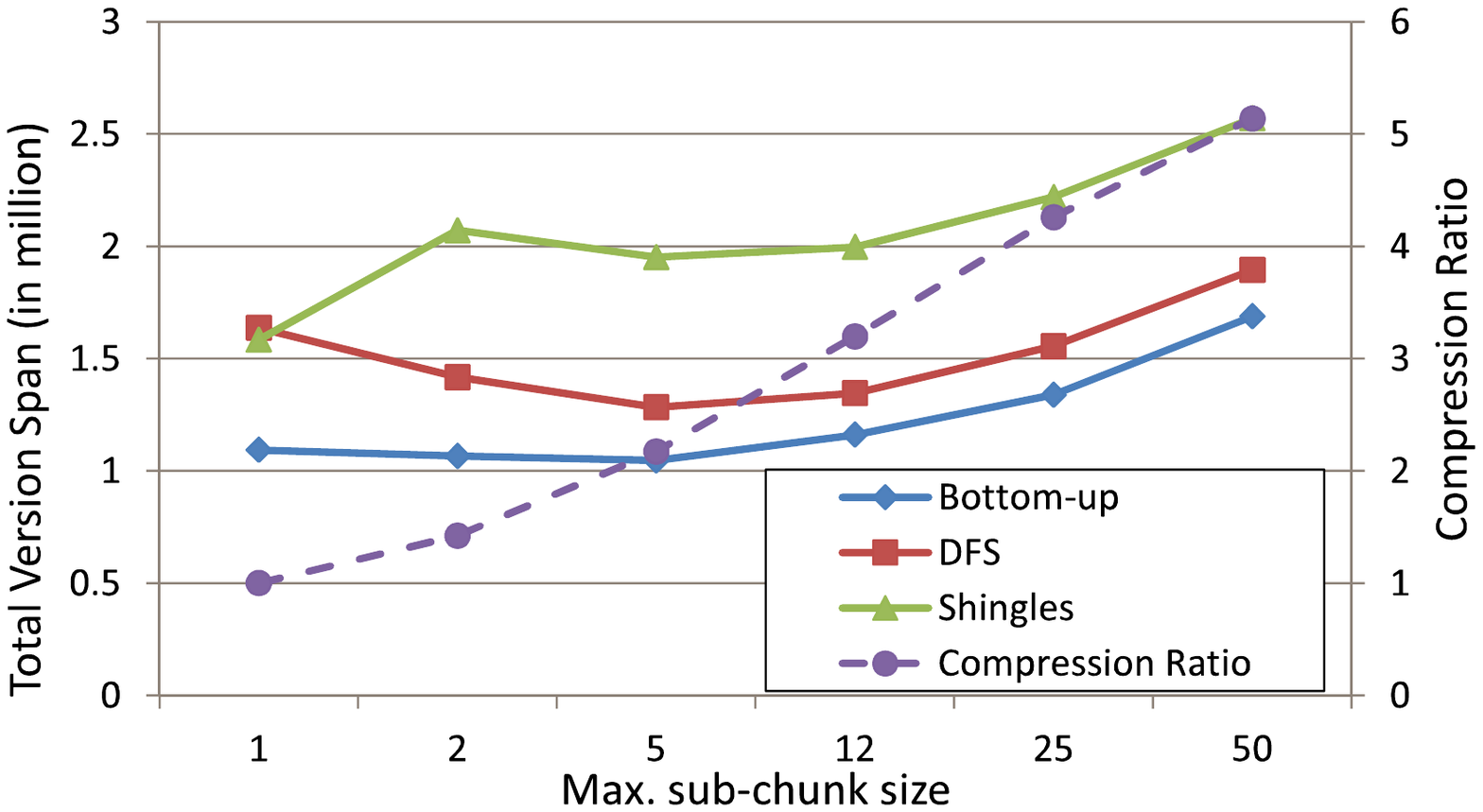}} &
\subfloat[\small Dataset D0, $P_d = 1\%$\label{subfig:d01}]{\includegraphics[trim=6cm 6.5cm 5cm 6cm, width=2.3in]{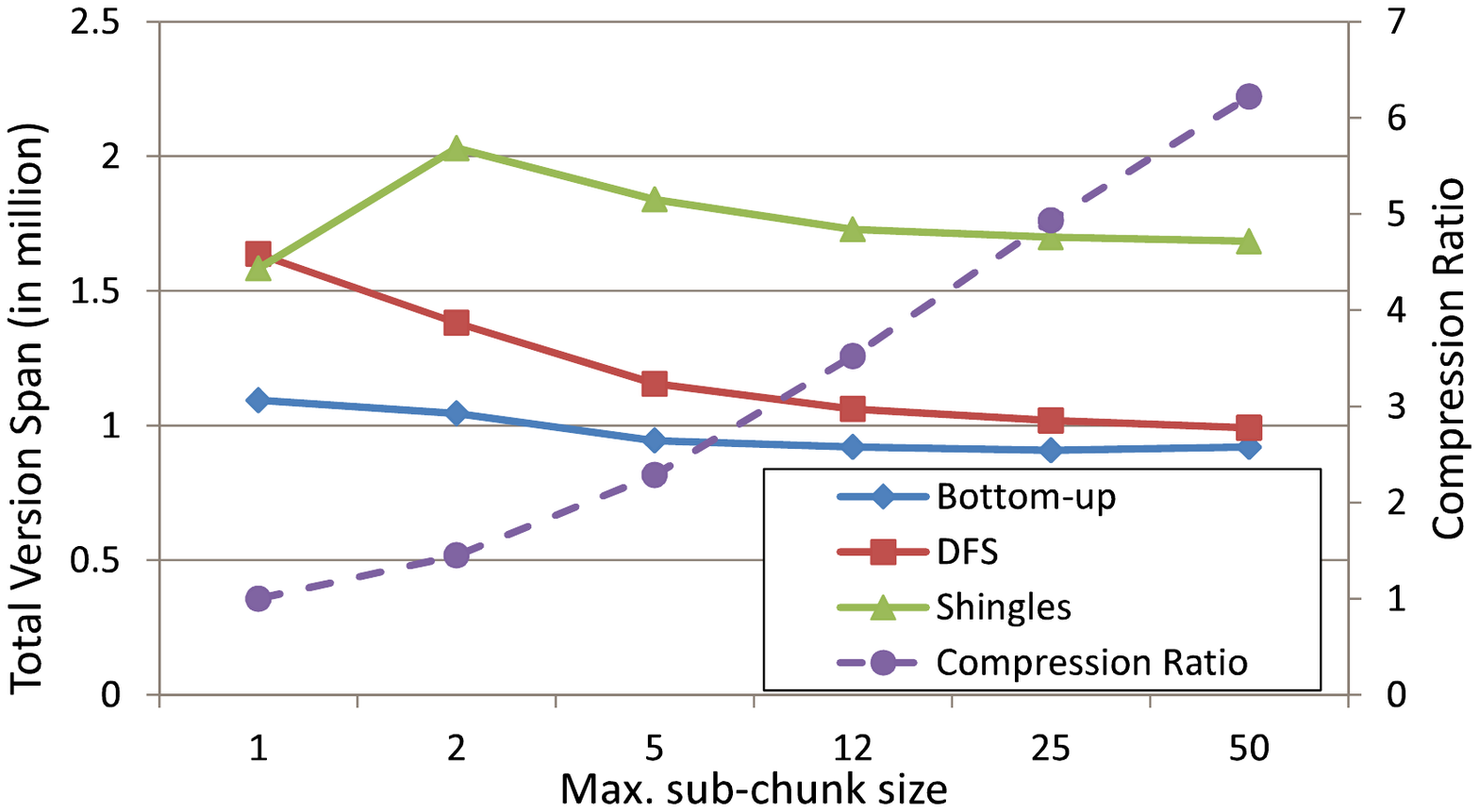}} 
\end{tabular}
\caption[]{\small Partitioning quality and compression ratios as sub-chunk size is varied for different algorithms}
\label{fig:compression-partitioning}
\end{figure*}

We now attempt to understand the performance of the partitioning algorithms on the compressed representation (Fig.~\ref{fig:compression-partitioning}). The degree of compression in the datasets is affected by two factors: (i) the number of records or the size of the sub-chunk, (ii) the amount of relative difference introduced between records due to updates. We simulate the second factor by generating the datasets such that when a record is updated, the amount of change w.r.t to the parent record is limited by a certain percentage, denoted by $P_d$. For a given version tree, we generate three datasets by limiting the change to 10$\%$, 5$\%$ and 1$\%$. For each such dataset, we vary the sizes of the sub-chunks (X-axis) and measure the total version span (Y-axis) at each sub-chunk value. We also plot the compression ratio (parallel Y-axis) of the dataset at every value of sub-chunk size. There are two factors that affect the total version span: (1) {\bf Sub-chunk size}: As the number of records in each sub-chunk increases, the total version span increases due to a decrease in the number of records fetched per chunk. (2) {\bf Compression Ratio}: Compressing the sub-chunks brings down the total number of chunks required to store the records. As a result, with increasing compression ratio the total version span is also expected to decrease. Note that we do not compare \deltat against these techniques as it is not possible to perform compression of records across multiple versions.

We observe that across all datasets, \bt has the best performance in terms of total version span. As $P_d$ decreases, we note that the total version span for same sub-chunk values decreases across all partitioning techniques and across all datasets. For example consider dataset C0, Fig.~\ref{subfig:c010}, Fig.~\ref{subfig:c05} and Fig.~\ref{subfig:c01}; the total version span at max sub-chunk size 50 decreases steadily with $P_d$ across all the partitioning techniques. This is because Factor 2 outperforms Factor 1 stated above and results in an overall decrease in total version span. However if we just consider Fig.~\ref{subfig:c010}, we observe an increase in total version span with max sub-chunk size which can be attributed to Factor 1 which is dominant here. But as we increase the degree of compression in Fig.~\ref{subfig:c05}, the effect of Factor 2 helps in reducing the effect of Factor 1, resulting in an overall reduction in total version span compared to the previous figure. Finally in Fig.~\ref{subfig:c01}, Factor 2 dominates over Factor 1 as the total version span now decreases with an increase in max sub-chunk size. This behavior was observed for Dataset D0 and other datasets as well (not plotted). However this is not true for Dataset A which is a linear chain as opposed to a branched tree like in the previous case. This is because Factor 2 has a more dominant role over Factor 1 due to the compression ratios which is better for dataset A compared to the other datasets.

\subsection{Query Processing Performance}\label{ssec:qp-perf}

\begin{figure*}[t]
\centering
\begin{tabular}{@{}ccc@{}}
\subfloat[\scriptsize Dataset A0, Q1 performance.~\subc: 4075.68s\label{subfig:a-q1}]{\includegraphics[trim=1cm 9cm 1cm 10cm,width=2.3in]{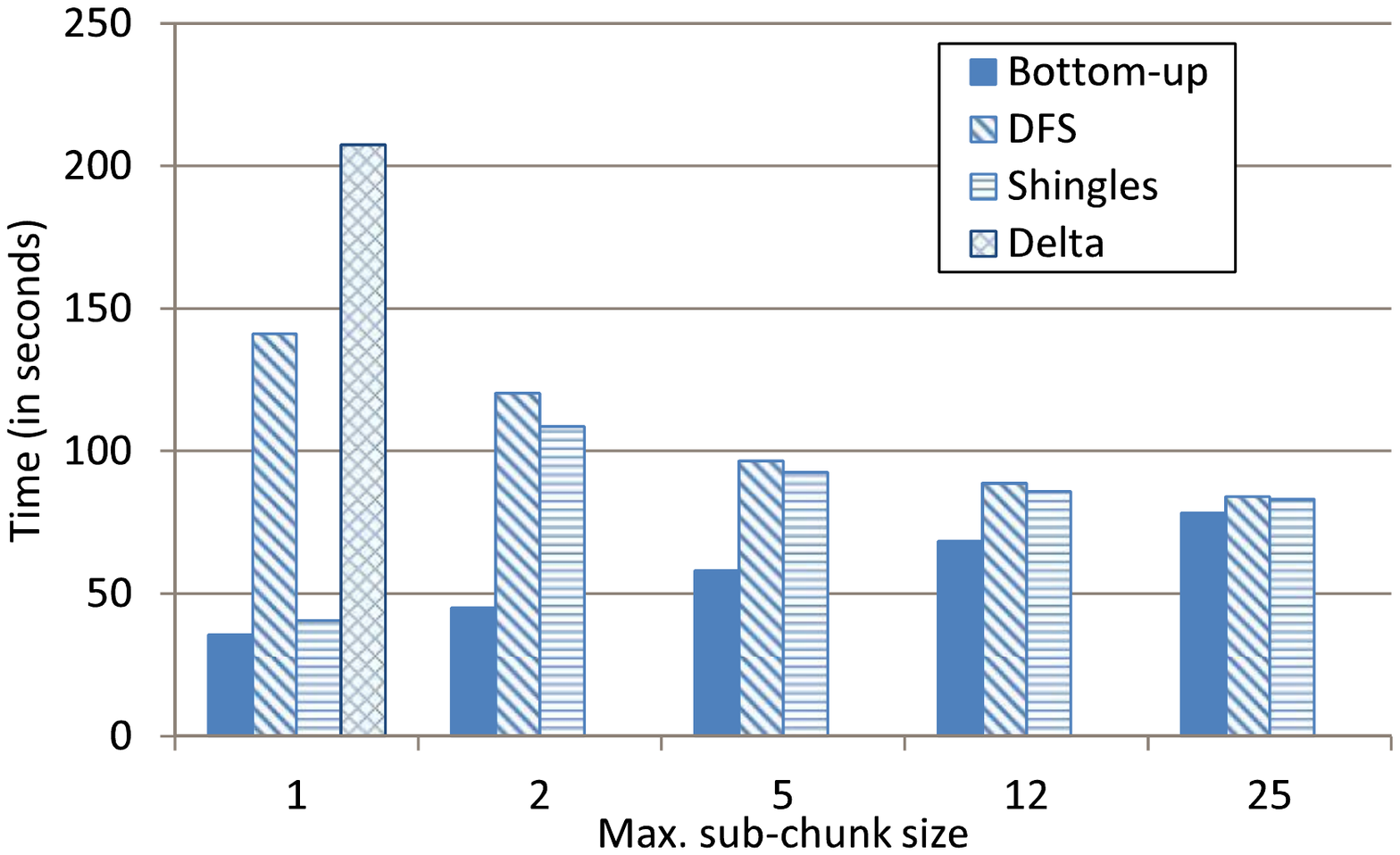}} & 
\subfloat[\scriptsize Dataset A0, Q2 performance.~\subc: 132.42s\label{subfig:a-q2}]{\includegraphics[trim=1cm 9cm 1cm 10cm,width=2.3in]{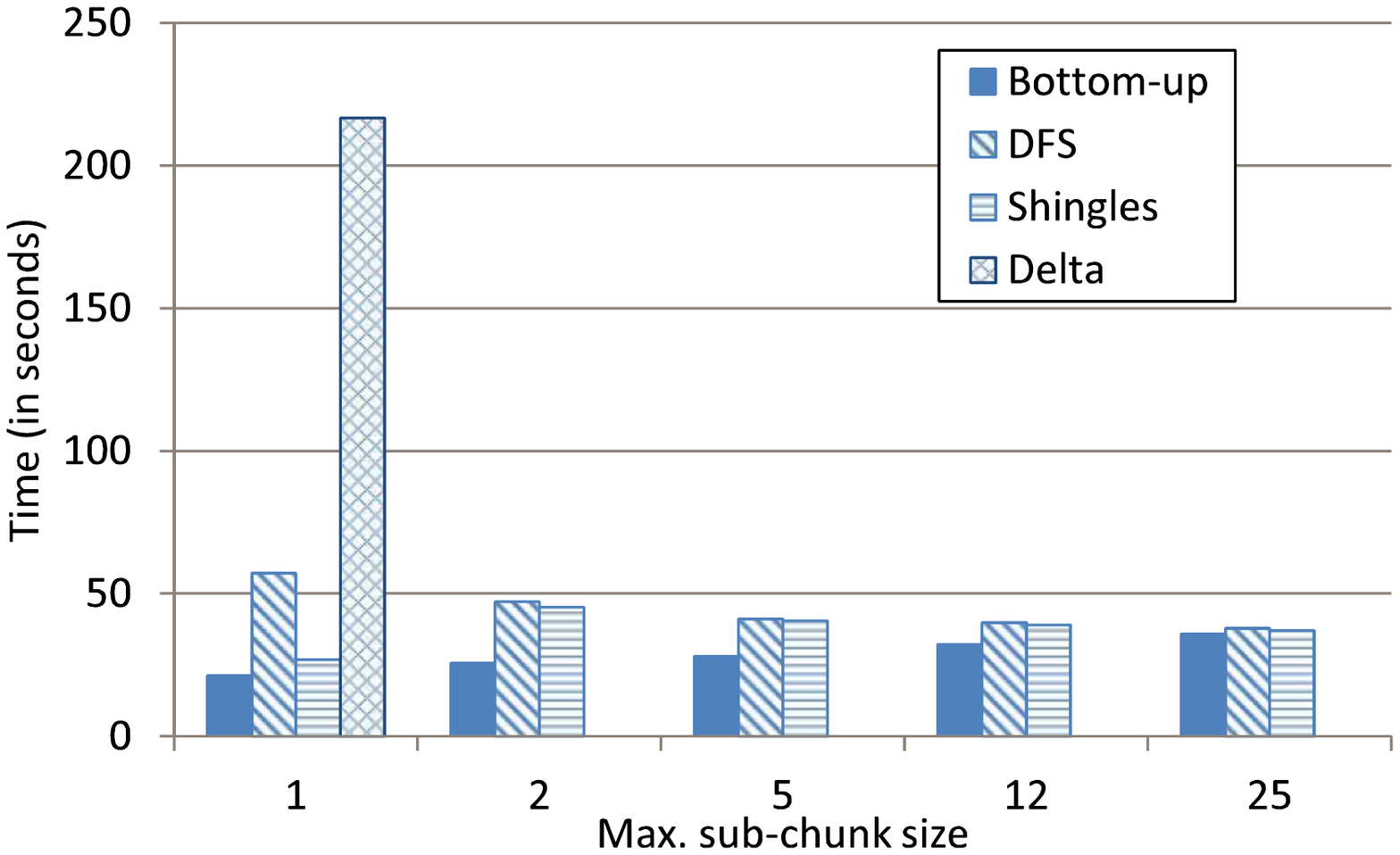}} &
\subfloat[\scriptsize Dataset A0, Q3 performance.~\subc: 0.0058s\label{subfig:a-q3}]{\includegraphics[trim=1cm 9cm 1cm 10cm,width=2.3in]{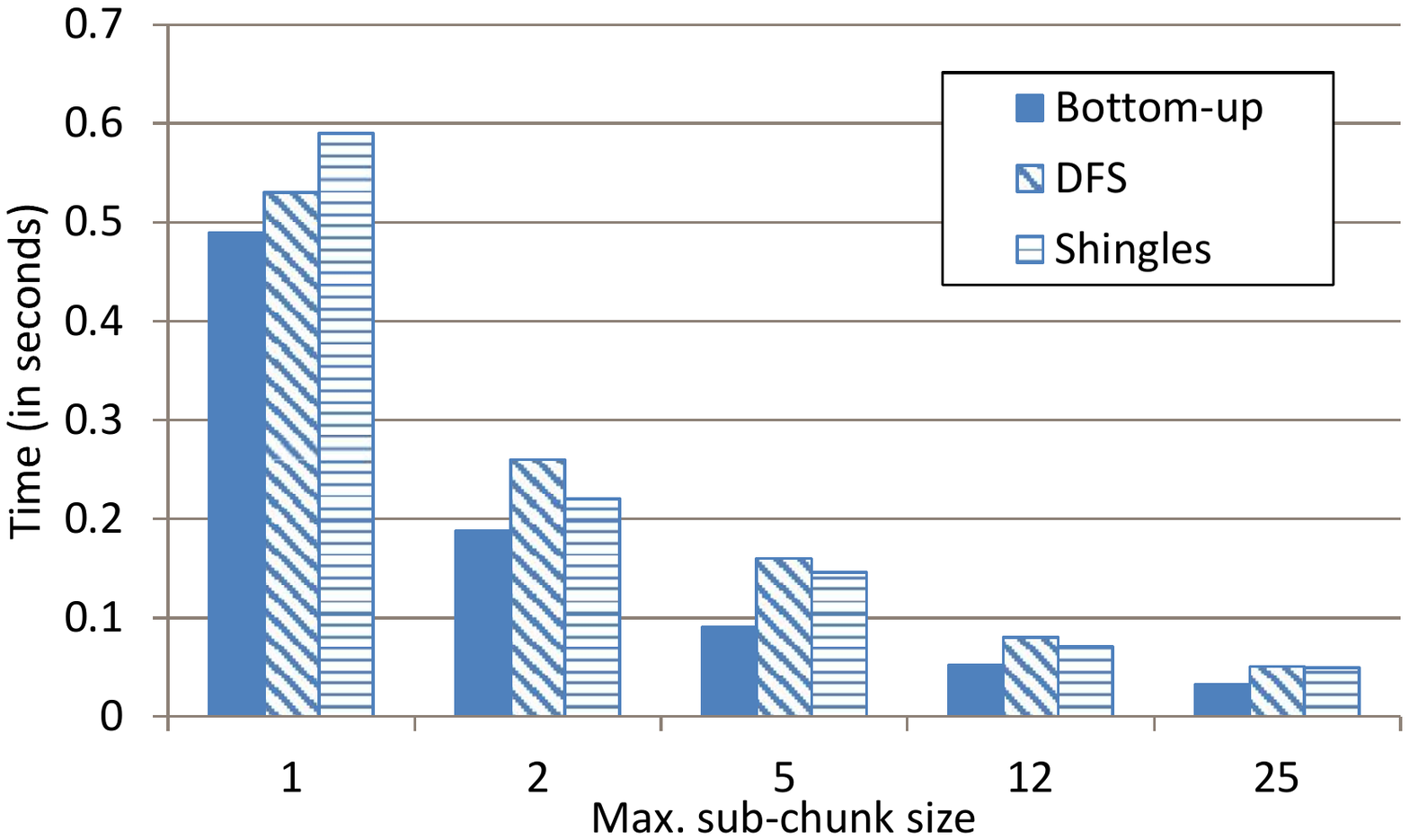}} \\[-4pt]
\subfloat[\scriptsize Dataset C0, Q1 performance.~\subc: 406.17s\label{subfig:c0-q1}]{\includegraphics[trim=1cm 9cm 1cm 9cm,width=2.3in]{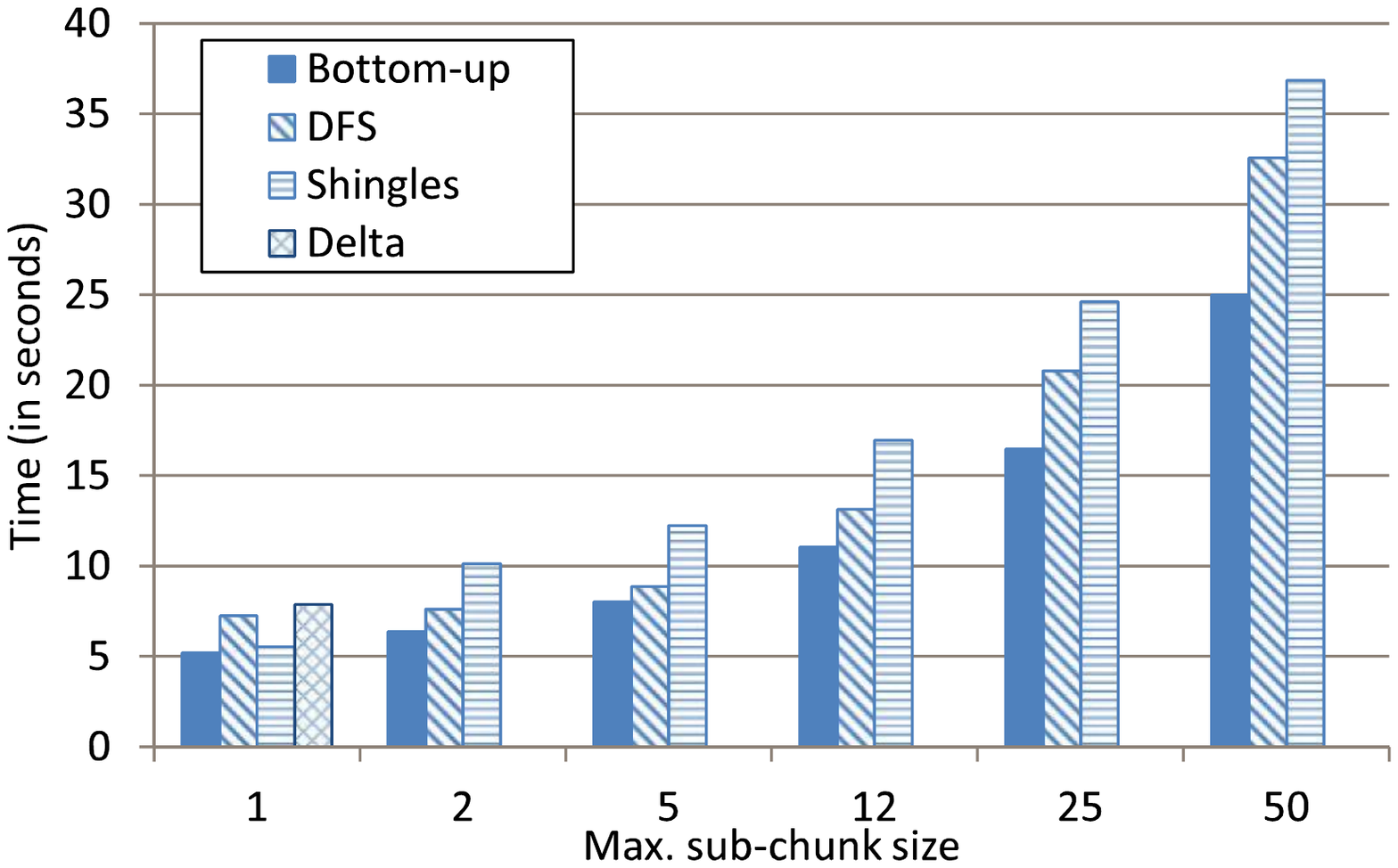}} & 
\subfloat[\scriptsize Dataset C0, Q2 performance.~\subc: 107.23s\label{subfig:c0-q2}]{\includegraphics[trim=1cm 9cm 1cm 9cm,width=2.3in]{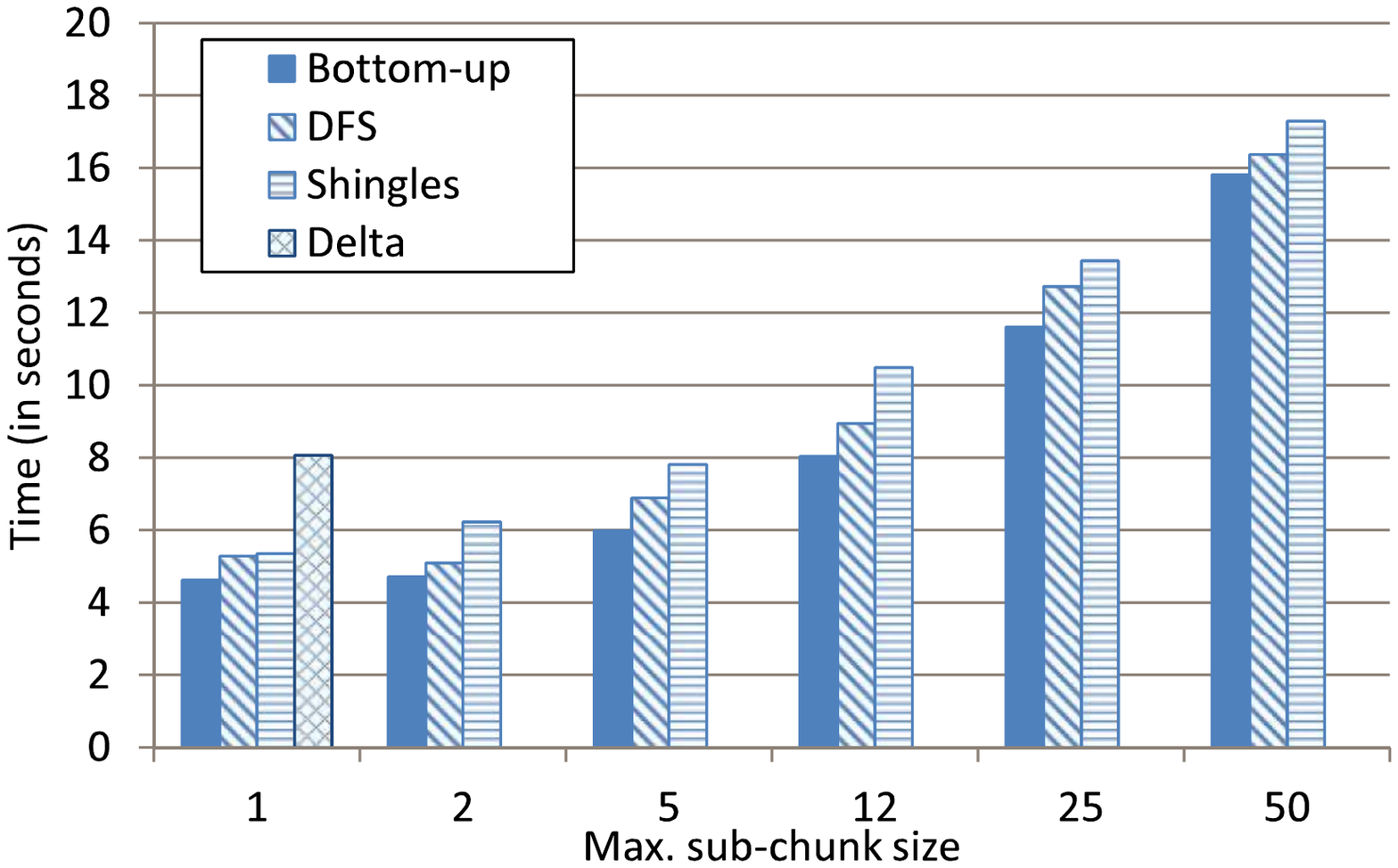}} &
\subfloat[\scriptsize Dataset C0, Q3 performance.~\subc: 0.0325s\label{subfig:c0-q3}]{\includegraphics[trim=1cm 9cm 1cm 9cm,width=2.3in]{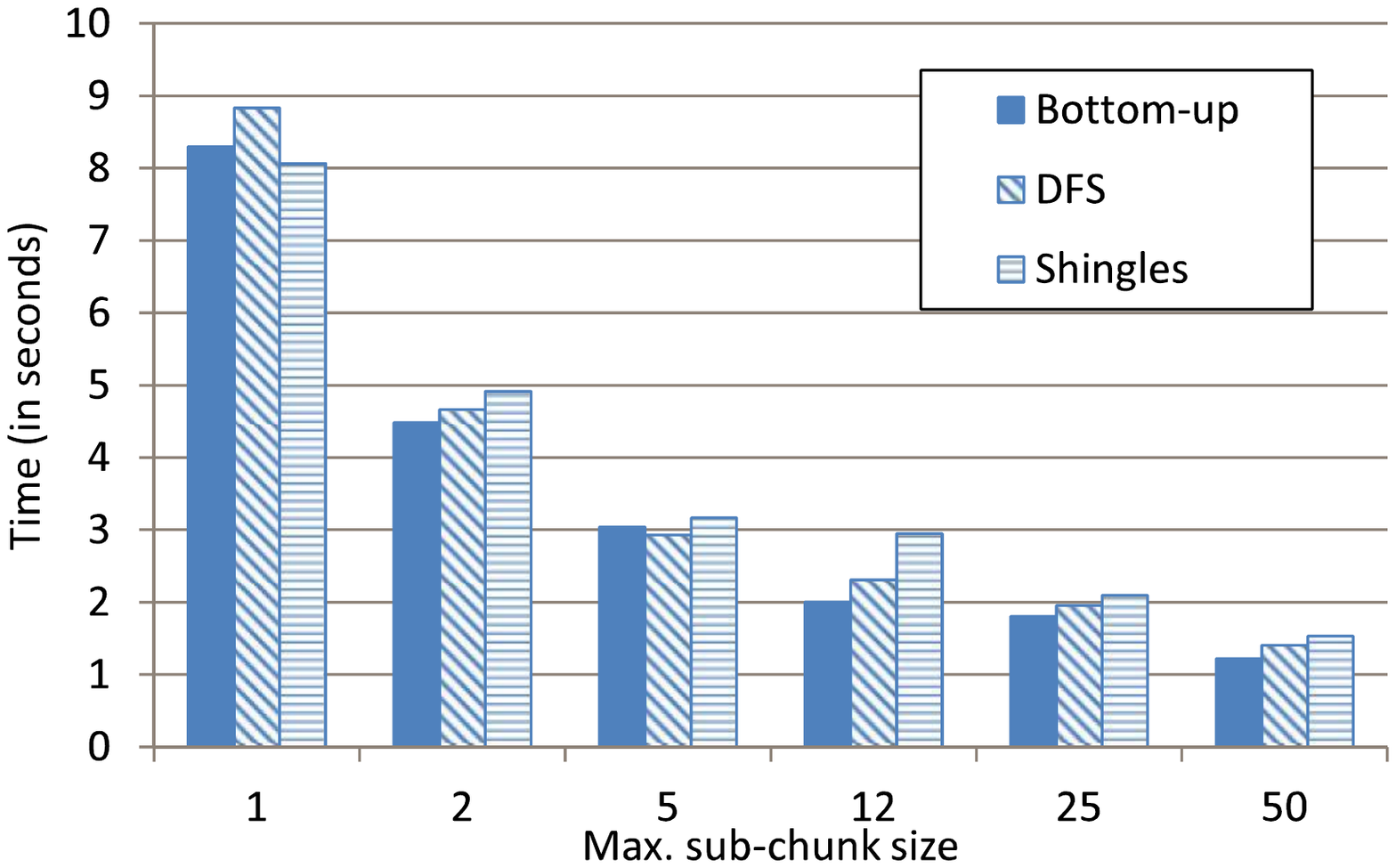}}
\end{tabular}
\vspace{-10pt}
\caption[]{\small Query Processing Performance}
\vspace{-10pt}
\label{fig:compression-qp}
\end{figure*}

In the following experiments (Fig.~\ref{fig:compression-qp}), we evaluate the query processing performance of \bt, \dfs, \sh and \deltat for three types of queries, namely, 1) Full Version Retrieval (Q1), 2) Partial Version Retrieval (Q2) and, 3) Record Evolution (Q3) on two different datasets. In all of these experiments we vary the max sub-chunk size from 1 to 50 and measure the total time taken to execute each of these queries against a randomly generated workload. 
Since intra-record compression is a limitation for \deltat, we restrict the \deltat experiment only to when the sub-chunk size is 1. We observe that \bt outperforms \dfs, \sh  and \deltat in terms of the query performance for Q1 and Q2; the performance curve of Q2 is similar to that of Q1 as partial version span is loosely proportional to full version span. Also note that time taken by \deltat for Q2 is greater than Q1. This is because in the worst-case the full version is first reconstructed and then the required records are filtered.

Recall that we fetch all the records corresponding to a primary key for Q3. Therefore we observe that storage representations with increasing sub-chunk sizes lead to better query processing performances for Q3. For \deltat, we need to reconstruct all the versions that and then filter out the required records which renders execution of Q3 impractical.
We also report the query performance of \subc technique against the caption of each query for each dataset. Although the full and partial version retrieval queries performs the worst for \subc, it outperforms all other techniques for record evolution query.

\subsection{Scalability of \ds}\label{ssec:scalability}

To demonstrate scalability of \ds, we ran a series of experiments where we doubled the cluster size starting at 1 up to 16, and then approximately double the amount of data by doubling the number of versions. We used two datasets specifically for this experiment, whose 16-node configurations were as follows: (a) {\bf Dataset G}: total size of the unique records = 255 GB, with 10k versions having $\approx$ 50K records each (version size: $\sim$275 GB, total size: 2.6 TB), (c) {\bf Dataset H}: size of the unique records = 280 GB, with 2k versions having $approx$ 100K records each (version size: $\sim$2.86 GB, total size: 5.76 TB). We partition the records using \bt approach. At each cluster configuration, we measure the full version retrieval times (partial version retrieval times showed similar behavior) and the record evolution times. As Fig.~\ref{fig:scalability} shows, \ds exhibits good {\em weak} scalability, and is able to handle appropriate larger datasets with larger clusters; the increased query times are largely attributable to increased version or key spans. We also note that \ds currently processes the retrieved chunks sequentially while constructing the query result and cannot benefit from the increased parallelism; we are working on parallelizing the entire end-to-end process, which will result in further improvements in the query latencies.

\begin{figure}[t]
\centering
\scriptsize
\begin{tabular}{|l|K{6mm}|K{5mm}|K{5mm}|K{5mm}|K{5mm}|K{5mm}|K{5mm}|}
  \hline
      {Query Worload} & Dataset & \multicolumn{6}{c|}{\# nodes in cluster} \\ \cline{3-8}
     {Avg. Version Span} & &1 & 2 & 4 & 8  & 12 & 16 \\ \hline \hline
     Q1 (in secs.) &  G& 7.35 & 7.95 & 8.99 & 10.49 & 10.97 & 11.39 \\ \cline{3-8}
     Avg. version span &  & 507.99 & 559.49 & 622.88 & 702.92 & 710.24 & 702.21 \\ \hline
   Q3 (in secs.) & G & 0.35 & 0.48 & 0.49 & 0.46 & 0.63 & 0.48 \\ \cline{3-8}
   Avg. key span &  & 21 & 32 & 34 & 33 & 46 & 34 \\ \hline \hline
   Q1 (in secs.) & H & 61.83 & 63.24 & 64.38 & 73.71 & 74.30 & 78.86 \\ \cline{3-8}
     Avg. version span & &400.24 & 436.48 & 451.20 & 554.92 & 561.60 & 594.92 \\ \hline
   Q3 (in secs.) & H &0.98 & 1.33 & 2.29 & 2.38 & 2.69 & 3.05 \\ \cline{3-8}
   Avg. key span & & 6 & 9 & 16 & 18 & 21 & 24 \\ \hline

\end{tabular}
\caption[]{Scalability Experiments}
\vspace{-12pt}
\label{fig:scalability}
\end{figure}

\subsection{Online Partitioning}\label{ssec:qp-online}

In this experiment (Fig.~\ref{fig:online-part}), we measure the performance of the online partitioning algorithm under different batch sizes for two datasets using the \bt partitioning technique. To measure the partitioning quality at a given point, we compute the ratio of the total version span obtained by online partitioning using that batch size, to that obtained by running an offline version of \bt for the same number of versions. Overall, even with small batch sizes, we observe reasonable penalties, with the partitioning quality improving with an increase in batch size. Thus, online partitioning without repartitioning, combined with a full repartitioning periodically, presents a pragmatic approach to handling updates.

\begin{figure}[t]
\scriptsize
\subfloat[Dataset B1\label{subtab:online-b1}]{
\begin{tabular}{|l||K{3mm}|K{3mm}|K{3mm}|K{4mm}|}
  \hline
      {Batch} & \multicolumn{4}{c|}{\# of versions} \\ \cline{2-5}
     {Size} & 250 & 500 & 750 & 1001 \\ \hline \hline
     125 & 1.13 & 1.36 & 1.52 & 1.63 \\ \hline
   250 & 1.00 & 1.12 & 1.23 & 1.32 \\ \hline
  500 & - & 1.00 & - & 1.10\\ \hline
\end{tabular}
}
\subfloat[Dataset C1\label{subtab:online-c1}]{
\begin{tabular}{|l||K{4mm}|K{4mm}|K{4mm}|K{5mm}|}
  \hline
      {Batch} & \multicolumn{4}{c|}{\# of versions} \\ \cline{2-5}
     {Size} & 2500 & 5000 & 7500 & 10001 \\ \hline \hline
     1250 & 1.04 & 1.05 & 1.06 & 1.08 \\ \hline
   2500 & 1.00 & 1.004 & 1.001 & 1.018 \\ \hline
  5000 & - & 1.00 & - & 1.005\\ \hline
\end{tabular}
}
\caption[]{Online Partitioning Performance}
\vspace{-12pt}
\label{fig:online-part}
\end{figure}

\section{Related Work}

Most cloud-based database systems including key-value stores primarily focus on providing efficient support for storing and retrieving data at the record level. Some of them provide support for additional features such as range queries~\cite{ChangDGHWBCFG06, PirzadehTPH12}; however it would be difficult for them to support range queries on versioned datasets in the absence of special indexes (Section~\ref{ssec:tradeoff}). Although there is no full-fledged support for managing multiple versions of the same record in these existing systems, there is some discussion about providing support for some {\em naive} form of versioning using the existing APIs in these systems. For example,~\cite{couchdb-blog, mongodb-blog} describe how to implement versioning features in Couchbase and MongoDB. The techniques described are similar and advocate storing previous versions of the record in a separate shadow {\em collection} before overwriting it with the updated value. A version number property (an \texttt{int32} called \texttt{\_version}) is added to the document to keep record of different versions. A downside of this approach as described is that records cannot be updated in batches and older versions are more expensive to retrieve. Moreover it is not clear if they support compressing multiple versions of the same record together.

\eat{
These cloud-based systems also use partitioning, in most cases, using a hash function, to distribute data across multiple nodes (or partitions). 
Although the problem of partitioning for minimizing query span has been addressed previously~\cite{Kumar13} by mapping to a hypergraph partitioning problem, the problem setting was quite different. 
The hypergraph constructed had data items or tuples as vertices and queries as hyperedges. 
In our setting, each record would map to a node in the hypergraph and a retrieval query would map to a hyperedge. 
Given the typical number of records in a dataset, the hypergraph would become very large, making their approach infeasible.
}

There has been significant work on workload-aware partitioning in recent years~\cite{Kumar13,PavloCZ12,CurinoZJM10,KumarQDK14}, with several of those approaches mapping the problem to a hypergraph partitioning problem with data items (records) as vertices and queries as hyperedges. 
Conceptually, the problem we address is identical, with the query workload defined by the verion retrieval queries. However, the sizes of the hyperedges for us are very large (since a version may contain millions of records) and those prior algorithms
(which implicitly assume small hyperedges) cannot be used. Another key difference is that, our algorithms exploit the inherent structure in the version graph.

There has been prior work on providing versioning support and compactly storing graph and XML data. \cite{buneman2004} proposed an archiving technique where all versions of the data are merged into one hierarchy. An element appearing in multiple versions is stored only once along with a timestamp. The hierarchical data and the resulting archive is represented in XML format which enables use of an XML compressor for compressing the archive. It was not, however, a full-fledged version control system representing an arbitrarily graph of versions; rather it focused on algorithms for compactly encoding a linear chain of versions. Moreover it does not provide support for range queries or record provenance queries that we support in our system. Finally their technique is not designed to work in a distributed setting that is essential for handling datasets that do not fit in a single machine.

There is extensive work on the temporal databases literature~\cite{Bolour92, SnodgrassA85, salzberg1999comparison} that manages a linear chain of versions and support version retrieval at a specific point in time. There, a specific version of a record/tuple is associated with a time interval, whereas in versioned databases, it is associated with a set of version-ids. This seemingly small difference leads to fundamentally different challenges -- e.g., whereas one could use an interval tree for indexing intervals optimally (e.g., to find all timestamps where a record is alive), doing the same for ``sets" is considered nearly impossible~\cite{hellerstein97}. An experimental evaluation in DEX~\cite{dex17} reveals that the techniques developed for linear chains~\cite{buneman2004} do not extend to branched version graphs.
There also has been prior work on compressing XML data~\cite{Liefke00} and providing update and versioning schemes for XML through an edit-based schema~\cite{chienTZ01}. \cite{chien2006toit} devises alternate storage schemes and provides support for complex queries on versioned XML documents. However the proposed techniques are single-node disk-based algorithms and are not designed to work in a distributed setting. There is work on developing a framework for incorporating temporal reasoning into RDF and studying the interplay between timestamp and snapshot semantics in temporal RDF graphs~\cite{gutierrezHV07}. \cite{snapshot2013} present an approach for managing historical graph data for large information networks, and for executing snapshot retrieval queries on them. 

Several version control systems geared towards handling different types of datasets have been recently developed, for unstructured files~\cite{BhattacherjeeCH15}, relational databases~\cite{MaddoxGEMPD16,orpheus}, arrays~\cite{MiaoLDD16}. Our work can be seen as exploring a different design point in that space, with a focus on storing versions of a collection of semi-structured or unstructured records in the cloud and supporting efficient key-based access to them.
\cite{orpheus} in particular proposes a partitioning scheme for minimizing version retrieval time by grouping versions into partitions and replicating records across them given a storage budget. Our approach of optimizing version retrieval cost does not consider replication of records across partitions. Further they do not support record compression grouped by keys to compress the data and then partitioning the compressed dataset.

Our approach is related to file content deduplication by {\em indexing} that employs hashes (or signatures) for identifying similar blocks of data~\cite{Broder1997, Quinlan2002}. The technique works by first identifying similar chunks of data across files or documents by computing a set of fingerprints for each chunk and then comparing the number of common fingerprints to assess the similarity. These fingerprints are then used to build indexes. Thus each block of chunk is stored once and each document can be represented by a collection of signatures. However these techniques do not involve any sort of partitioning.

\section{Conclusion}
We designed and built a system for managing a large number of versions and branches of a collection of keyed records in a distributed hosted environment, and systematically analyzed
the different trade-offs therein. Our work is motivated by the popularity of key-value stores for storing large collections of keyed records or documents, the increasing trend towards maintaining histories of {\em all} changes that have been made to the data at a fine granularity, and the desire to collaboratively analyze and simultaneously modify or 
transform datasets. We showed that simple baseline approaches to adapting a key-value store to add versioning functionality suffer from serious limitations, and proposed a flexible and tunable framework intended to be used a layer on top of any key-value store. We also designed several novel algorithms for solving the key optimization problem of partitioning records into chunks. Through an extensive set of experiments, we validated our claims, design decisions, and our partitioning algorithms. 
%
%
%
%
In future work, we plan to develop more sophisticated online algorithms that effectively re-partition the records as new records are committed into the system. We also aim to explore the effect of replication as it reduces the cost of version reconstruction but increases the cost of storing the versions.

{
\bibliographystyle{abbrv}
\balance
\bibliography{references}
}

\end{document}